\documentclass{aa}  
\usepackage{graphicx}
\usepackage{txfonts}
\usepackage{xurl}
\usepackage{changes}
\usepackage{natbib}
\usepackage{pdflscape}
\usepackage{xcolor}	
\usepackage{longtable}
\usepackage{subcaption}
\usepackage{tabularx}

\usepackage{hyperref}
\hypersetup{colorlinks=true,linkcolor=blue,citecolor=blue,filecolor=blue,urlcolor=blue}
\begin{document} 

\title{SIEGE III: The formation of dense stellar clusters in sub-parsec resolution cosmological simulations with
individual star feedback}

\titlerunning{Resolving star clusters in cosmological simulations}

\authorrunning{Calura et al.}
\author{
  F.~Calura \inst{1},  R.~Pascale\inst{1}, O. Agertz\inst{2}, E. Andersson\inst{3}, E. Lacchin\inst{4,5,6}, A. Lupi\inst{7,8}, M. Meneghetti\inst{1},
  C. Nipoti\inst{9}, A. Ragagnin\inst{1}, J. Rosdahl\inst{10}, E. Vanzella\inst{1}, E. Vesperini\inst{11}, A. Zanella\inst{12,1} 
}

\institute{
INAF-Osservatorio di Astrofisica e Scienza dello Spazio di Bologna, Via Gobetti 93/3, 40129 Bologna, Italy
\and Lund Observatory, Division of Astrophysics, Department of Physics, Lund University, Box 43, SE-221 00 Lund, Sweden
\and Department of Astrophysics, American Museum of Natural History, New York, NY 10024, USA
\and Dipartimento di Fisica e Astronomia “Galileo Galilei”, Università di Padova, Vicolo dell’Osservatorio 3, 35122 Padova, Italy
\and INFN – Padova, Via Marzolo 8, 35131 Padova, Italy
\and Institut fuer Theoretische Astrophysik, ZAH, Universitaet Heidelberg, Albert-Ueberle-Straße 2, 69120 Heidelberg, Germany
\and DiSAT, Universit\`a degli Studi dell’Insubria, via Valleggio 11, I-22100 Como, Italy
\and INFN, Sezione di Milano-Bicocca, Piazza della Scienza 3, I-20126 Milano, Italy
\and Dipartimento di Fisica e Astronomia “Augusto Righi”, Università di Bologna, Via Gobetti 93/2, 40129 Bologna, Italy
\and Universit\'e Claude Bernard Lyon 1, CRAL UMR5574, ENS de Lyon, CNRS, Villeurbanne F-69622, France
\and Department of Astronomy, Indiana University, Bloomington, IN 47401, USA
\and INAF – Osservatorio Astronomico di Padova, vicolo dell’Osservatorio 5, 35122 Padova, Italy
}

\date{  Received \today ;  Accepted   }

\abstract
{Star clusters stand at the crossroads between galaxies and single stars.
  Resolving the formation of star clusters in cosmological simulations represents an ambitious and challenging goal,
  since modelling their internal properties requires very high resolution. 
This paper is the third of a series within the SImulating the Environment where Globular clusters Emerged (SIEGE) project, where
we conduct zoom-in cosmological simulations with sub-parsec resolution that include the feedback of individual stars, aimed to model
the formation of star clusters in high-redshift proto-galaxies.
We investigate the role of three fundamental quantities in shaping the intrinsic properties of star clusters, i. e., 
i) pre-supernova stellar feedback (continuous or instantaneous ejection of mass and energy through stellar winds);
ii) star formation efficiency, defined as the fraction of gas converted into stars per
freefall time, for which we test 2 different values ($\epsilon_{\rm ff}=0.1$ and 1), 
and iii) stellar initial mass function (IMF, standard vs top-heavy).
All our simulations are run down to $z=10.5$, which is sufficient
for investigating some structural properties of the emerging clumps and clusters. 
Among the analysed quantities, the gas properties are primarily sensitive to the feedback prescriptions. A 
gentle and continuous feedback from stellar winds originates a complex,
filamentary cold gas distribution, opposite to explosive feedback, causing smoother clumps.  
The prescription for a continuous, low-intensity feedback, along with the adoption of $\epsilon_{\rm ff}=1$, also 
produces star clusters with maximum stellar density values 
up to  $10^4 M_{\odot}$ pc$^{-2}$, in good agreement with 
the surface density-size relation observed in local young star clusters (YSCs).
Therefore, a realistic stellar wind description and 
a high star formation effiency are the key ingredients that
allow us to achieve realistic star clusters characterised by properties comparable to those of local YSCs. 
In contrast, the other models produce too diffuse clusters, in particular the one with a top-heavy IMF.}


\maketitle
%
\section{Introduction} 
Resolving the formation of star clusters stands out as one of the most ambitious goals in galaxy formation models.
Stellar clusters are key to the formation of stars, as increasing evidence suggests that most stars (if not all) are born in various forms
of aggregates, such as groups, clusters, or hierarchies of these systems \citep{lad03,rod20}.
On the other hand, star clusters strongly affect their surrounding 
environment through their strong mass and energy outputs, by driving super-bubbles of hot gas that trigger galactic outflows 
\citep{ten05,bik18,lev21,orr22}. 
Moreover, a large fraction of stars in 
various galactic components, such as the Milky Way halo and discs, originated from dissolved star clusters \citep{bic01,wie88,kru19,rei23}. 
Star clusters are often regarded as self-standing entities whose formation,
evolution, and possible dissolution were viewed in relative 
isolation, considering their host galaxy as the background source of a passive tidal field (e.g., \citealt{sol21,ves21,lac24}).

Galaxy formation models have frequently described star clusters as macroscopic particles 
(often called 'star particles'),  
typically representing simple stellar populations and ignoring their 
underlying substructure (e. g., \citealt{age13,sti13,hop18,fel23}).  
In current efforts to model the formation and evolution of star clusters within the context of galaxies and cosmology, it has
become increasingly clear that this separation between galaxy and star cluster evolution is no longer tenable. As these two are intricately linked, understanding the co-evolution of galaxies and their embedded star clusters has become of crucial importance. 

In recent times, thanks to the first observational studies of the progenitors of globular clusters (GCs) at high redshift
\citep{van17a,van17b,cal21,bou21,mow22,mpas23,sen24}, 
star clusters have regained significant interest also from a cosmological perspective. 
These discoveries motivated several attempts to account for the presence of star clusters in
galaxy formation models. 

Hydrodynamic simulations performed in a cosmological framework are valuable tools 
to model realistically several fundamental physical processes,  such as the gravitational collapse,
radiative cooling and star formation (SF). 
They are sometimes based on ‘zoom-in’ techniques, in which a low-resolution dark matter (DM)- only
simulation is first run, starting from initial conditions computed self-consistently with the adopted cosmology and up to a certain
epoch of interest. When a DM halo presents relevant features for the purpose of the study (e. g., it has a suitable virial mass value),
higher resolution simulations are then run, centered on that halo and including also the baryonic physics. 
In most cases, current state-of-the-art simulations reach a spatial resolution that is inadequate to resolve the
formation of star clusters in early galaxies. 
A suitable spatial resolution for this purpose needs to be at a sub-pc scale, sufficient to capture rapid, small-scale key
processes such as tidal shocks, as well as the turbulent nature of SF \citep{ren13,ren20}.

Only a few studies so far have a suitable resolution to investigate the physical conditions in which star clusters originate;
however, they suffer from severe limitations. 
These works resolve the formation of star clusters, although their sizes are generally larger than the observed
ones due to limited resolution \citep{ma20}. 
In some very high-resolution simulations of this type, the stellar component is modelled
by means of stellar particles, aimed to represent entire stellar populations \citep{kim16,gar23}. 
However, the increase in resolution is accompanied by decreasing reliability of 
the sub-grid description of stars with particles. 
In paricular, \cite{rev16} showed how below a critical mass particle of $10^3$ M$_\odot$,
only a direct, star-by-star sampling of the IMF provides a realistic description of the stellar component in hydrodynamic simulations 
(see also \citealt{eme19}). 
Generating individual stars is necessary when the quantity of gas available for SF in a cell is sufficient for a few
stars only, which becomes increasingly frequent with sub-pc resolution and gas densities of $n\sim10^3-10^5$ particles cm$^{-3}$,
typical of molecular cloud cores.  
Several simulations describe isolated systems, such as star clusters or galactic systems, in which
the stellar component is modelled by means of individual stars that release mass, energy 
and heavy element in their surroundings \citep{eme19,and20,lah20,wal20,gut21,hir21,hu23,den24}. 
In these works, the initial and boundary conditions are 
idealized or simplified and do not represent the physical complexity of real star-forming systems and their environment,
for which the full description of a cosmological framework is required. \\
The physics of stellar feedback (including stellar winds, supernovae and various 
radiative processes) sets the amount of mass and energy released by stars into the interstellar medium (ISM) 
and is a natural consequence 
of stellar evolution. Within star-forming dense clouds, the timescale on which feedback acts
on the system is debated, in particular it is yet unclear whether the most fundamental properties
of the system are determined before or after supernova (SN) explosions (e. g., \citealt{age13,gee16,kru19,che20,and24}).
Another major, unanswered question concerns the role of the star formation efficiency (SFE) and, in particular,
how it regulates the formation of bound star clusters  \citep{bau07,gru21,fuk21,pol24}. 
Finally, the number of sources mainly contributing to stellar feedback, i.e. massive stars,  
is strongly dependent on the IMF, whose shape and
evolution is still largely unknown. 
The roles of these aspects altogether on star cluster formation and
their complex interplay have never been addressed 
in cosmological simulations. 

Within a project aimed at SImulating the Environment where Globular clusters Emerged (SIEGE), in a previous work we developed 
a zoom-in cosmological simulation of a high-redshift dwarf galaxy at sub-pc resolution, including feedback from individual stars (\citealt{cal22}, hereinafter FC22). 
The model is aimed to describe a strongly lensed star-forming complex observed at $z=6.14$  
which includes a few star clusters \citep{van19,cal21,mes24}. 

In their first work, the stellar systems in the simulations of \cite{cal22} presented very diffuse stellar clumps,
with properties marginally similar to those  of high-redshift star forming complexes but significantly different
from those of stellar clusters.  
Building on the model presented in \cite{cal22}, in the present paper we investigate further this issue and 
consider how different prescriptions for stellar feedback, SFE and stellar initial mass function
impact on star cluster properties, such as density and size.

This paper is organized as follows.
In Section 2 we describe the main features of the simulations and the model assumptions.
In Sect. 3 we present our results, whereas in Sect. 4 we draw our conclusions.
The flat cosmological model adopted throughout this paper has 
matter density parameter $\Omega_m$ = 0.276
and Hubble constant $H_0$ = 70.3 km s$^{-1}$ Mpc$^{-1}$ \citep{omo19}.

\section{Simulations setup}
The target of our zoom-in simulations is a DM halo of mass $\sim 4 \times 10^{10} M_{\odot}$ at $z=6.14$,
aimed to represent the host of multiple star-forming clumps, 
in a system that is the theoretical analogue of the D1-T1 stellar complex, detected through strong gravitational
lensing and containing dense stellar systems that qualify as globular cluster precursors \citep{van19,cal21}. 
The cosmological initial conditions are computed as described in FC22. 
The choice of a comoving volume of 5 Mpc h$^{-1}$  is key for achieving sub-pc resolution throughout the duration of
our simulations, i. e. from z = 100  to z = 10.5.
The zoom-in region is defined by means of dark matter-only simulations and a multi-step method, in which
we incrementally increase the resolution at intermediate steps (\citealt{fia17,lup19}; for further deatils, see FC22).
We use the adaptive mesh refinement RAMSES code \citep{tey02}, that solves the 
Euler equations with a second-order Godunov method and adopt an HLLC Riemann solver.\\
In cells eligible for SF, 
instead of spawing star particles (representing entire stellar populations that sample the full 
initial mass function, IMF), we stochastically sample the IMF to generate individual stars \citep{sor17,cal22}.   
Both the stellar and DM components are modelled through collisionless particles, whose 
trajectories are computed by means of a Particle-Mesh solver. 
For the DM, the maximum mass resolution is of 200 M$_{\odot}$.\\
We adopt a quasi-lagrangian refinement strategy, based on the number of particles present in a cell.
Specifically, a cell is refined when its mass exceeds $8 \times m_{\rm SPH}$, where $m_{\rm SPH}= $32M$_{\odot}$.
This allows us to reach a maximum 
physical resolution of 0.2 pc in the densest regions at z = 10.5, corresponding to maximum refinement level lmax = 21.\\
Despite the fact that star clusters are widely known to be
fundamentally collisional systems (e. g., \citealt{spi87}), in this work we neglect the effects of
gravitational collisions between stars.  
Our choice is justified by recent studies showing that in the long-term evolution of GCs, tidal heating dominates over internal
processes, such as stellar collisions \citep{car22}. Addressing the relative roles of the cosmological 
tidal field and internal processes in the dynamical evolution of star clusters is the focus of an on-going project \citep{ves24}. 

As in FC22, we do not consider any pre-enrichment from population III stars, whose
feedback is though to affect the properties of mini-halos at very high redshift and enrich the surrounding intergalactic medium
with metals (e. g., \cite{jeo17,kle23}. 
Investigating their effects is a significant endeavour, as it requires exploring various parameters, including the stellar yields, 
(e. g., \citealt{heg10}), variations in the 
metallicity transition and initial mass function  \citep{jeo17}. 
A study of the effects of individual pop III stars will be the subject of future work.

We present a set of simulations in which we test the effects
of various quantities (feedback, stellar initial mass function, star formation efficiency)
on the properties of the stellar systems formed in the computational box.  
A summary of the features of the simulations run in this work is presented in Table \ref{table_sim},
useful to illustrate the investigated parameters. 

\begin{table*}
    \small
  \caption{Summary of the simulations run in this work} 
  \begin{tabular}{l|ccccc|l}
  \hline
  Model name    &   pre-SN feedback    & IMF   &  SF efficiency    &  Maximum resolution$^1$           & Stellar Mass   &   Other properties        \\
                 &   (ejection)                   &       &   $\epsilon_{\rm ff} $ &                        &  at $z=10.5$   &    (common to all models)  \\    
\hline
 FC22             &    instantaneous     &  Kroupa (2001)  &   0.1       &  0.2 pc            & $1.7\times10^6$M$_{\odot}$     &      Individual star mass range:     \\
 Winds, SFE=0.1    &    continuous          &  Kroupa (2001)  &   0.1       &  0.2 pc          & $3.5\times10^6$M$_{\odot}$      &    $>$1.5 $M_{\odot}$        \\
 Winds, SFE=1.0-LR &    continuous          &  Kroupa (2001)  &   1.0       &  0.8 pc          & $7.6\times10^6$M$_{\odot}$     &    Massive stars range:          \\
 Winds, SFE=1.0    &    continuous          &  Kroupa (2001)  &   1.0       &  0.2 pc          & $5.8\times10^6$M$_{\odot}$     &    8-40 $M_{\odot}$  \\
 Winds, THIMF      &    continuous          &  $\frac{dn}{dlog m}\propto$ const.  &  0.1 & 0.2 pc & $1.7\times10^6$M$_{\odot}$  &    M$_{\rm ej}^2$=0.9 m$_{\rm ini}$\\
                  &                        &                 &                   &               &                           &  Stellar lifetimes: C15 \\
\hline	      
\end{tabular}
1: Computed at z=15.5, corresponding approximately to the onset of star formation.\\
2: Total mass ejected by massive stars at the end of their lives\\
\label{table_sim}

\end{table*}
\begin{figure*}
\center
\includegraphics[width=18.cm,height=12.75cm]{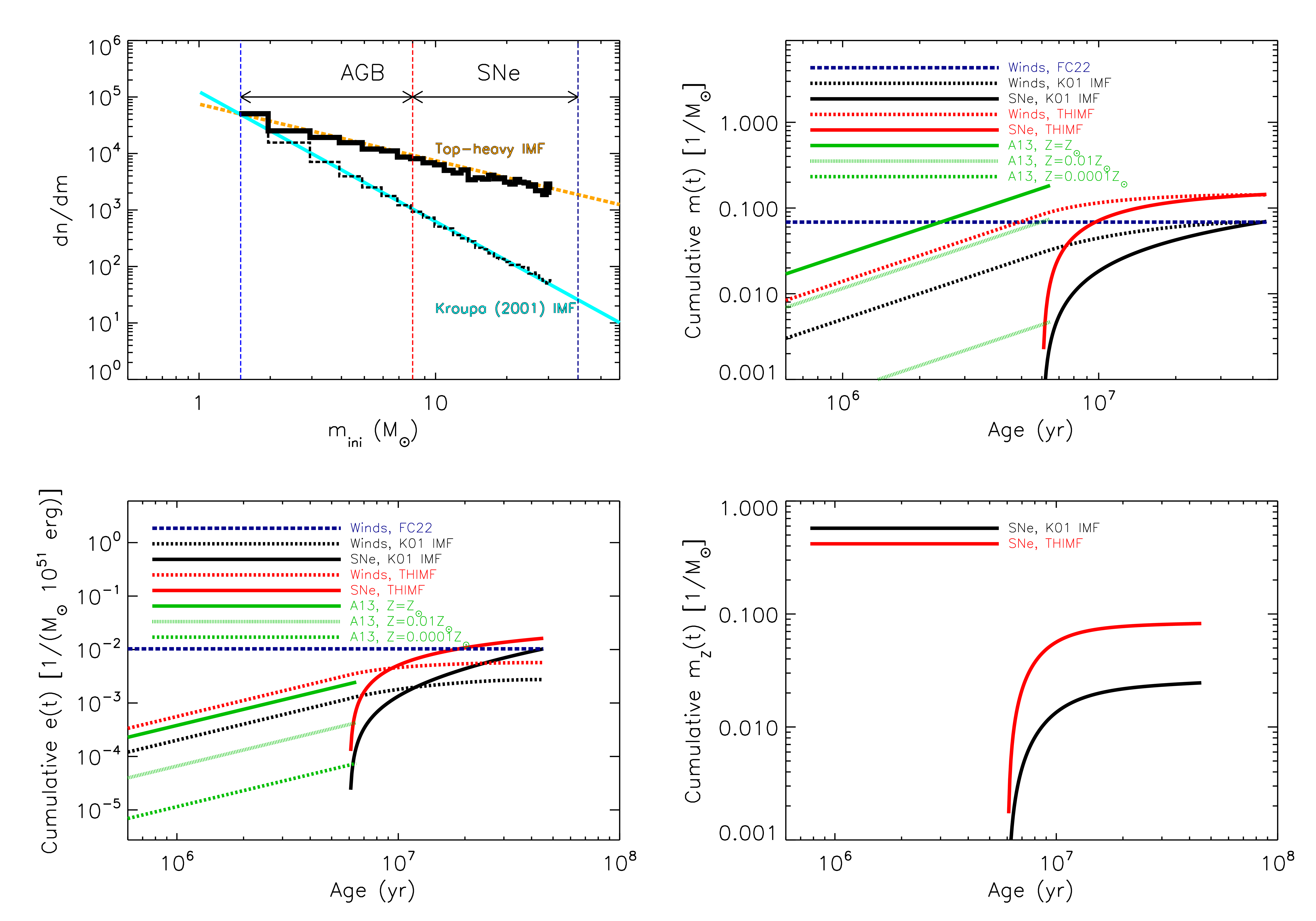}
\caption{Top-left: stellar IMF computed in this work for a standard (black solid line)
  and a top-heavy (black dashed line)
  IMF compared to the analytic ones (cyan solid line: \cite{kro01}; orange dashed line: THIMF).
  The vertical blue, red and dark-blue dashed lines indicate the minimum mass for individual
  stars, for massive stars and the maximum mass for SN progenitors, respectively.
  The top-right is the cumulative mass per unit stellar mass ejected by a stellar
  population for a \cite{kro01} (black lines) and top-heavy IMF (red lines)
  for stellar winds (dashed lines) and SNe (solid lines), whereas the blue dashed line represents the 
  prescriptions for stellar winds ejecta adopted in FC22.
  The solid, dashed and dotted green straight lines are the metallicity-dependent cumulative mass
    injected by stellar winds in the model of \cite{age13} and stopping at $t=$6.5 Myr, corresponding to
  the wind duration in their model. 
  The bottom-left panel is the energy per unit mass and in units of  $10^{51}$ erg
  ejected by stellar winds and SNe in a stellar population, with same
  line types as in the top-right panel. The bottom-right panel is the specific cumulative mass
  in the form of heavy elements for a standard (black line) and top-heavy (red line) IMF.
  } 
\label{fig_stellar}
\end{figure*}

\begin{figure*}
\center
\includegraphics[width=17.cm,height=12.75cm]{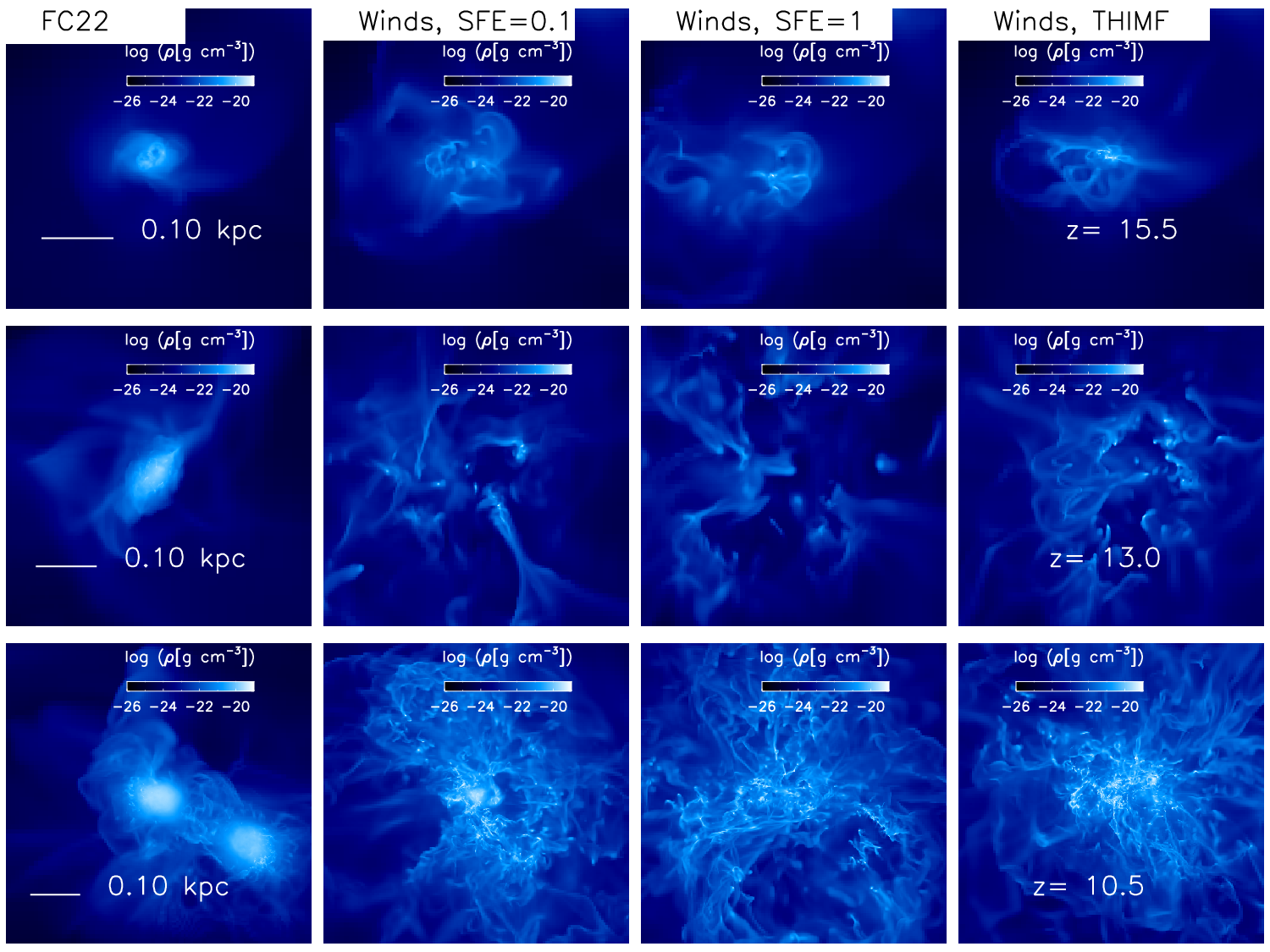}
\caption{Slice gas density maps in the x-y plane in the four models presented in Table \ref{table_sim} 
  at different redshifts.
  The maps describe the density distribution in the central region of the simulations in the 
  FC22 (first column from left), 'Winds, SFE=0.1' (second), 
  'Winds, SFE=1.0' (third) and 'Winds, THIMF' models.  
  The maps are at $z=15.5$ (upper row), $z=13$ (middle row) and  $z=10.5$ (bottom row). 
  The horizontal white solid lines shown in the left-hand column-panels indicate the physical scale. } 
\label{fig_map_gas}
\end{figure*}

\begin{figure}
\center
\includegraphics[width=8.cm,height=19.cm]{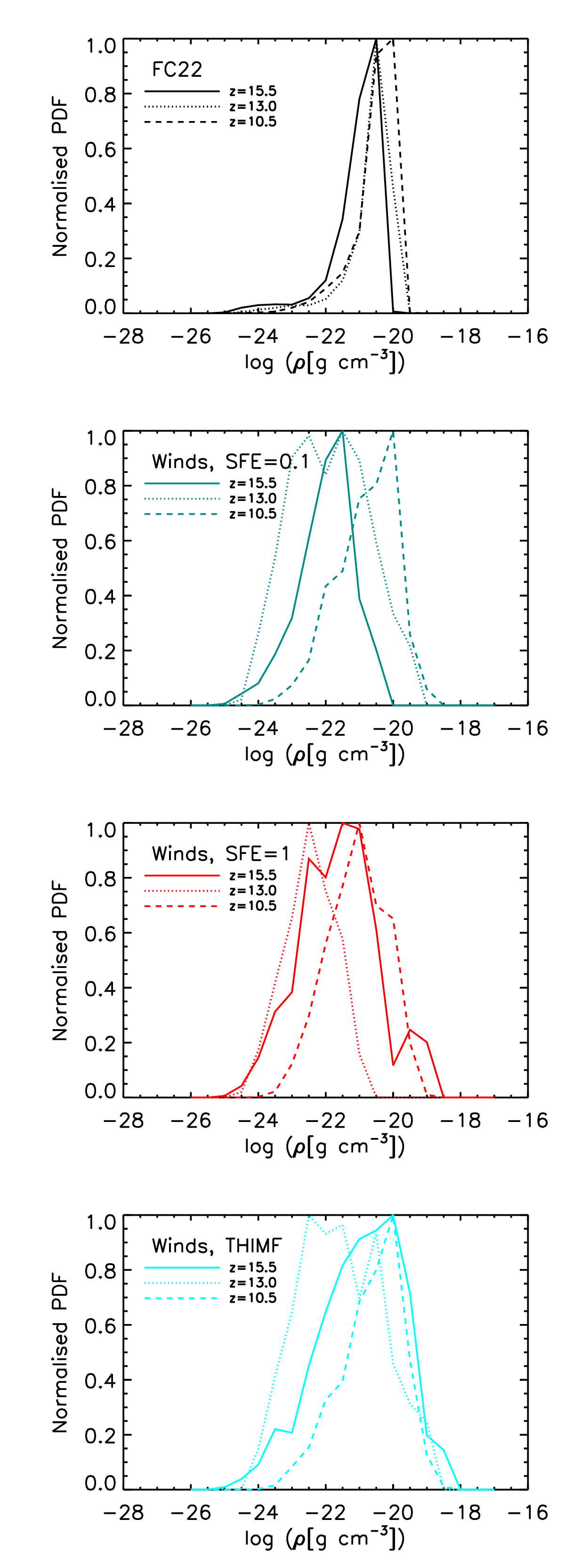}
\caption{Redshift evolution of the normalised (with respect to the maximum)  
  probability distribution function of the gas density in the same
  central region of our simulations as in Fig. \ref{fig_map_gas}.
  From top to bottom: gas density PDF in the FC22 (first panel), 'Winds, SFE=0.1' (second), 
  'Winds, SFE=1.0' (third) and 'Winds, THIMF' (fourth) models at $z=15.5$ (solid lines),  
  $z=13$ (dotted lines) and $z=10$ (dashed lines). } 
\label{fig_pdf_gas}
\end{figure}

\subsection{Generation of individual stars}
\label{sec_imf}
In RAMSES \citep{tey02}, the rate at which gas is converted into new stars is expressed by the \cite{sch59} law  
\begin{equation}
  \dot{\rho}_{\rm *} = \frac{\rho}{t_*}, 
\end{equation}
where $\rho$ and ${\rho}_{\rm *}$ are the gas and stellar density, respectively, and $t_*$ represents the
star formation timescale.
This quantity is proportional to the local freefall time $t_{\rm ff}=\sqrt{3~\pi/32~G \rho}$ and is expressed as 
\begin{equation}
  t_{\rm *}=  t_{\rm ff}/\epsilon_{\rm ff},  
\end{equation}
where $\epsilon_{\rm ff}$ is the star formation efficiency per freefall time, a fundamental quantity related to the
intensity of SF and a free parameter. 
In the local Universe, SF is known to be an inefficient process, with typical
$\epsilon_{\rm ff}$ values of one or a few percent on the scales of individual giant molecular clouds  
(GMCs, e. g., \citealt{mye86,mur11,fed12,gru19}).
However, while most observations of GMCs find very 
similar median values, different surveys suggest a significant spread in $\epsilon_{\rm ff}$
(\citealt{mye86,mur11,eva14,lee16,vut16,uto18,gru19,gri19}; see also our Discussion). 
Deep observations of giant molecular clouds in lensed, star-forming galaxies support higher 
values for $\epsilon_{\rm ff}$ at high redshift \citep{des23},
and observations of local starbursts as well \citep{fis22}, where the physical conditions of the star-forming gas 
are plausibly more similar to the ones of early galaxies \citep{hec98,pet09,sil15}.  

In simulations, $\epsilon_{\rm ff}$ can span 3 order of magnitudes, 
from a typical value of one or a few percent in galaxy formation
simulations aimed to describe the large-scale properties of
local galaxies (e. g., \citealt{gho24,seg24}),  
up to 100 $\%$ in works that aim to model globular cluster formation \citep{li18,bro22,lah23}.

In the model of FC22 we assumed $\epsilon_{\rm ff}=0.1$, supported by previous results
from cosmological simulations of MW-sized galaxies, showing how such choice
is consistent with direct observations in molecular clouds and allows one to reproduce fundamentals scaling relations,
such as the Kennicutt-Schmidt, and a set of other observables \citep{age15}.
Later results of isolated MW-like systems suggest that models where 
star formation is regulated by stellar feedback require $\epsilon_{\rm ff}=0.1$ 
\citep{gri19}. 
In this work we test two different SFE values, i. e.  $\epsilon_{\rm ff}=0.1$ and $\epsilon_{\rm ff}=1.0$, 
and assess their effects on the formation of dense stellar aggregates. 

We assume that star formation can occur in cells with gas temperature $T<2 \times 10^4$ K. 
The total mass available for star formation in a cell is split into an array of individual stars using the following
method \citep{sor17}. 
We decompose the stellar mass spectrum into $N$ finite mass intervals. 
In each interval a mass fraction $f_i$ is defined, so that 
\begin{equation}
\sum_{i=1}^{N} f_{i} = 1. 
\end{equation}
In the $i-$th interval, the number of individual stars $n_i$ is sampled  
from a Poisson distribution, characterised by a probability $P_i$ given by
\begin{equation}
P_i(n_i)={\lambda_i^{n_i} \over n_i !} \exp({-\lambda_i}) \,
\label{poisson_law}
\end{equation} 
where the mean value $\lambda_i$ is 
\begin{equation}
\lambda_i= f_{i} {M \over m_i}. 
\label{poisson_param}
\end{equation}
In the equation above, $M$ is the total mass available for star formation
(allowing for no more than  90 \% of the gas in the cell is 
turned into stars), whereas   
$m_i$ is the average stellar mass in the $i-$th bin. We use this formalism to determine the number of individual stars produced in each bin,
except for the lowest mass bin, representing stars with mass below 1.5 M$_{\odot}$ and where star particles are spawned that collect all
the lowest-mass stars together (for further details, see FC22). 

To ensure an adequate representation of the IMF, we adopt $N=100$ linearly spaced mass bins. 
In each bin, the calculation of mass fraction $f_i$ requires the assumption of a stellar IMF. 
The IMF of the first stars is largely unknown and matter of intense debate (e. g., \citealt{lar98,glo05,cla11,hir14}),
as it depends on the interplay of various processes that include (proto-)stellar feedback, metallicity, accretion 
and gas fragmentation \citep{bro04,fer04,kle23}. 
In the early Universe, the different chemical composition of the cold gas and, in particular, the absence of heavy
elements may affect significantly the radiative cooling and cloud fragmentation, 
producing an overabundance of massive stars or the formation of extremely massive objects (e. g., \citealt{saf14,hir17,cho21}. 
Considering this substantial uncertainty, in this work we will test two different choices for the stellar IMF. 
In most simulations, we assume a \cite{kro01} (K01) IMF, very common in the local Universe and defined as: 

\begin{equation}
\frac{d~n}{d~\log~m} (m) =  \Big\{ \begin{array}{l l} 
                                      A \cdot m^{-0.3} & 
			     \qquad {\mathrm{if}} \; m < 0.5 \, \rm M_\odot \\
			              B \cdot m^{-1.3} &
			     \qquad {\mathrm{if}} \; m \ge 0.5 \, \rm M_\odot. \\
                                      \end{array}                                      
\end{equation}
The K01 IMF is defined in the mass range $0.1 \rm M_\odot \le m \le 100 \rm M_\odot$. \\
To account for the possible overabundance of massive stars in the early Universe, we also test the effects
of a top-heavy IMF (THIMF),
for which we consider the simple and convenient functional form 
\begin{equation}
\frac{d~n}{d~\log~m} (m) = \rm constant 
\end{equation}
(e. g., \citealt{jeo17}) and the same mass range as for the K01. 
The IMFs adopted in this paper are shown in the top-left panel of Fig. \ref{fig_stellar}.
Note that in this plot, the IMF is expressed as $\frac{d~n}{d~m} (m) = \frac{d~n}{d~\log~m} \frac{1}{m}$, that is a different 
conventional way to plot this function and in which the top-heavy IMF is $\frac{d~n}{d~m} \propto m^{-1}$
(in these units, the canonical \cite{sal55} form is $\propto m^{-2.35}$).
Here, we plot the IMFs in the mass range where individual stars are defined, i.e. between $\sim 1.5$ and $100~M_{\odot}$. 
To maximize the number of stars, the numerical IMFs (black solid and dashed lines) have been calculated
at the final times of our simulations.

When generating individual stars, the total stellar mass must  not exceed the mass available within the cell.
With the exception of a few very rare cases, this causes the truncation of both IMFs at $m \sim 30 \rm M_\odot$.
Our choice in favor of the conservation of mass 
causes the loss of a negligible number of massive stars from the total budget, without any significant effect on
stellar feedback.

\subsection{Stellar feedback prescriptions}
\label{sec_feed}
We assume that stars in the mass range $8 \rm M_\odot \le m \le 40 \rm M_\odot$ contribute to stellar feedback
through the release of mass and energy into the ISM in both the pre-SN and SN phases. 
We test different stellar feedback prescriptions and compare the present results with the ones obtained 
in the FC22 simulation, where, to model the two phases, all the mass and energy were returned by massive stars instantaneously in two
separate episodes, i. e. at their birth and at their death. 
Since in FC22 we considered 12 mass bins only for the IMF, with a consequent poorer sampling of the stellar mass range,
the previous simulations were re-run with the prescriptions described in Sect. \ref{sec_imf} for a \cite{kro01} IMF. 

In cosmological simulations where star clusters are not resolved,
pre-SN feedback has an important role in enhancing the overall effect of stellar feedback.
On galactic Scales (e. g., \citealt{age13,hop18})
momentum injection from stellar radiation, winds and SNe are generally comparable,
but SNe (in the post Sedov-Taylor phase) will dominate momentum unless the effect of infrared optical depth is strong \citep{age13}. 
However, a few pieces of evidence have suggested  that pre-supernova (SN) feedback
is fundamental in driving the evolution of young stellar 
clusters and their sorrounding environment \citep{hop10,dkru19},
with SNe explosions expected to play a minor role 
in key processes, such as the dispersion of gas in molecular clouds \citep{che22}. 
Moreover, other studies have shown how the cumulative energy delivered
by massive stars in the pre-SN phase is comparable to that released
by SN explosions (\citealt{cas75,ros14,cal15,fie16}). 
In light of these arguments, in all our simulations, it is crucial to have an adequate description of stellar feedback in the pre-SN phase of massive stars.
Through simplified prescriptions, we model the effects of stellar winds from massive stars. In the following, we describe the
basic ingredients of our model. 
In the 'Winds' models of this work, in the pre-SN phase and starting immediately after its birth,
each massive stars constantly restores mass and energy through a continuous stellar wind. 
The constant mass and energy return rates are $\dot{M}$ and $\dot{E}$, respectively, and
are proportional to the initial mass $m_{\rm ini}$, expressed as
\begin{equation}
  \dot{M} = \eta \frac{m_{\rm ini}}{\tau_m}
  \label{eq_mdot}
\end{equation}
and 
\begin{equation}
  \dot{E} = \frac{\dot{M} v_w^2}{2}.
  \label{eq_edot}
\end{equation}
In equations  \ref{eq_mdot} and \ref{eq_edot}, $\eta=0.45$ whereas $\tau_m$ is the stellar lifetime, for which we consider
the analytic 
fit of \cite{cai15} (C15) of the \cite{por98} results, whereas in eq. ~\ref{eq_edot} $v_w=2000$ km s$^{-1}$ is the assumed value for the terminal wind velocity
(e. g., \citealt{wea77,vin18}).

Our assumptions concerning the cumulative masses ejected by single stars in the wind phase are consistent with
  the predictions from stellar evolution models at solar metallicity \citep{ren17}. 

We do not assume any dependence of the mass return rate on stellar metallicity.
We are aware that this assumption is an oversimplifcation,
and is likely to lead us to overestimate the effects of stellar winds on the regulation of SF in star clusers (see below).

As in FC22, each massive star ends its life exploding as a SN, returning a fraction $\eta$ of its initial mass and 
an amount of thermal energy of $10^{51}$ erg. 

The spatial concentration of massive stars can be very large and this can cause extremely large gas temperatures,
causing excessively small timesteps in the simulation. To prevent such overheating, 
we set a maximum value for the temperature ($T_{\rm max}=10^8$ K) for the hot medium driven by stellar feedback.  
This value is of the order of the post-shock temperature of the medium energised by stellar winds  \citep{mac23}. \\

As in FC22, we adopt the native, metal-dependent implementation of radiative cooling of the RAMSES code, based on equilibrium-thermochemistry.
The cooling and heating rates of the gas
are computed as a function of the temperature, density, redshift, metallicity,
and the abundances of a set of primordial ion species that include ions of H and He.

To prevent numerical overcooling in high density regions of the ISM,
we adopt a delayed cooling scheme as described in FC22. This assumption consists 
in temporarily switching off cooling in
suitable cells, in which the feedback is released as thermal energy. 
Effective feedback is obtained by assuming that each massive star injects in its cell also 
an amount of 'non-thermal' energy, stored in a passive tracer variable and
ideally associated with an unresolved, generic turbulent energy. 
In the native implementation of this scheme \citep{tey13}, radiative cooling 
is switched off in each cell in which the local abundance of the 'non-thermal' passive tracer
is above a certain threshold; moreover, the latter is 
assumed to decay on a dissipation time-scale. 
As explained in FC22, without any knowledge of the appropriate values for these two quantities and considering also
that in the present simulations we model single stars, and not star particles representing entire simple stellar populations, 
we choose to constrain these two parameters empirically, starting from known recipes from the literature
tested at different regimes of mass and spatial resolution \citep{dub15}.   
By conducting low-resolution tests, we have confirmed that this approach accurately describes stellar feedback
and allows us to reproduce the stellar mass of the observed system.  
On empirical grounds, our choice enables an effective model which 
allows us to achieve the desired results, i.e. an efficient feedback at
our resolution and considering our ingredients. 

As for metal production, we adopt the same prescriptions as in FC22. 
The amount of metals ejected by 
each single star is described by an analytic formula that represents a polynomial fit to the
\cite{woo95} stellar yields
\begin{equation}
y_{\rm Z} = \Sigma_{k=1}^6 f_k ~ (m_{\rm ini}/\rm M_{\odot})^k
\label{eq_yield}
\end{equation}
with $f_k = [0.0108, -0.0026, 0.00026, -8.81~\times~10^{-6}$, $1.23~\times~10^{-7}$ and $-0.5791]$.

Our prescriptions for stellar feedback are summarised in Fig. ~\ref{fig_stellar}, where we show the cumulative
specific mass (top-right panel), energy (bottom-left panel) and heavy elements mass (bottom-right panel) returned by 
massive stars in a simple stellar population, calculated assuming a K01 and a top-heavy IMF. 
In our 'Winds' models, the cumulative mass returned by stellar winds and SNe after 30 Myr is equivalent.

Fig. ~\ref{fig_stellar} clearly shows how adopting a THIMF causes a stronger energy release. 
In such case, the cumulative total mass and specific energy returned by a simple stellar population is a factor $\sim 2$ larger than the one obtained with
a K01 IMF.
The 'FC22' model is the one characherised  by the largest amount of mass and energy deposited by a newly born stellar population. 
In fact, it is worth noting that even with a THIMF, the energy deposited soon after the birth, e. g. at $10^6$ yr,
is more than one order of magnitude smaller than the one of  the FC22 model.

On the other hand, the THIMF causes a  considerable enhancement of the cumulative metal fraction, up to a factor $\sim 3$ larger than the K01. 
This reflects the strong dependence of the metal yields (defined as the specific amount of mass returned by each star in the form of heavy
elements) on the initial mass in the prescriptions considered in this work (see FC22). 

It is worth noting that stellar mass-loss rates strongly depend on the metallicity and that in stellar
evolution models,  stellar winds from low-metallicity stars are much weaker during the main sequence phase
(e.g., \citealt{lim18}). 
In our simulations, a significant fraction of stars are born with zero
(or very low) metallicity (Ragagnin et al., in prep.), which, in principle, should have very
little effect on star formation. Our assumption of a metallicity-independent prescriptiond for stellar winds
leads to overestimating their feedback. This is shown also by the comparison of
our prescriptions adopting a K01 IMF and the analytical fits to the metal-dependent cumulative mass loss
and energy release estimated by \citealt{age13} (green solid, dashed and dotted lines in Fig. ~\ref{fig_stellar}),
computed from the the STARBURST99 code \citep{lei99}. We are aware that 
our adoption of  metallicity independent rates is an oversimplification that requires improvement.
In a forthcoming work, we plan to include metallicity-dependent wind rates (e. g., \citealt{dib11,den24}) in our models,
to study the effects of low-metallicty stellar winds on stellar cluster properties.  \\
Moreover, in general, considering  stellar winds as the primary feedback process and neglecting effects form
photoionisation radiation represents another oversimplification.
In fact, \cite{lan21b} showed that in dense clouds, turbulent mixing enhances
energy losses from the hot interior, and the efficiently-cooled, momentum-driven wind
bubbles are not expected to be dominant. In ``normal'' GMCs (with typical surface density
$\Sigma_{\rm gas} \sim 10^2$ M$_{\odot}$ pc$^{-2}$), 
SF regulation occurs mostly by photoionisation \citep{kim18}. \\
Finally, the progenitors of  asymptotic giant branch (AGB) stars are intermediate-mass stars (with a mass $m<8 \rm M_{\odot}$),
which return their ejecta 
instantaneously at the end of their life. 
For these stars, we adopt an analytic fit to the final-to-initial stellar mass relation derived by \cite{cum18}.

\section{Results}
In all our simulations, star formation begins at z = 15.95 which, for the adopted cosmology, corresponds 
to a cosmic time of 0.251 Gyr. 
In this section, we present our results in the form of slice or projected maps of some relevant quantities that
describe the properties of the gas and the stars, computed at three reference, equally spaced redshift values,
i. e. $z=15.5$, $z=13$ and $z=10.5$. 
The choice of running the simulations to $z=10.5$ is compliant with the parameter space
we aim to explore (see Tab. \ref{table_sim}) and
the considerable computational time required by our runs. 
Moreover, this cosmic time interval is sufficient to gain a clear insight into the properties
of the systems formed in our simulations, the differences between the models and the roles played by the most relevant
quantities.  
Most of the results presented in this work concern the central stellar clump, i. e., the
stellar aggregate that lies at the centre of the zoom-in region. 
A more detailed analysis of the general properties of the stellar clumps of the present simulations
will be presented in a future study (Pascale et al., in prep.).

\begin{figure*}
\center
\includegraphics[width=17.cm,height=12.75cm]{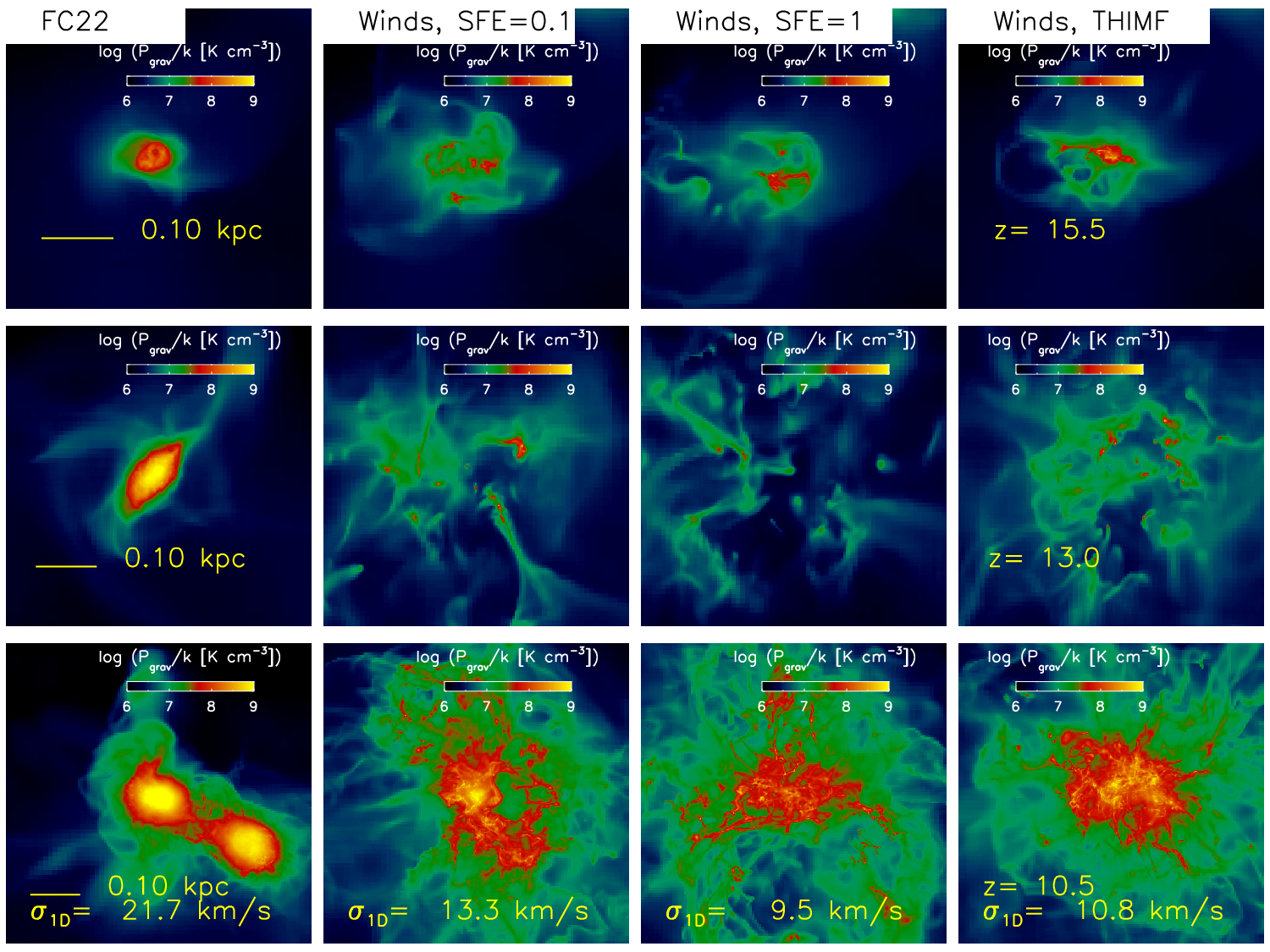}
\caption{Maps of the gravitational pressure $P_{\rm grav} = G\,\Sigma_{\rm gas}^2$ of the gas (see the text for details)  
  in the x-y plane and in the central region of the simulations in the 
  FC22 (first column from left), 'Winds, SFE=0.1' (second), 
  'Winds, SFE=1.0' (third) and 'Winds, THIMF' (fourth) models.  
  The maps are at $z=15.5$ (upper row), $z=13$ (middle row) and  $z=10.5$ (bottom row). 
  The horizontal white solid lines shown in the left-hand column-panels indicate the physical scale.
  In each panel we report also the 1D density-weighted velocity dispersion of the cold medium (with temperature $<200$ K)
  calculated as in Eq. ~\ref{eq_sigma}.} 
\label{fig_press_gas}
\end{figure*}

\begin{figure*}
\center
\includegraphics[width=18.cm,height=4.5cm]{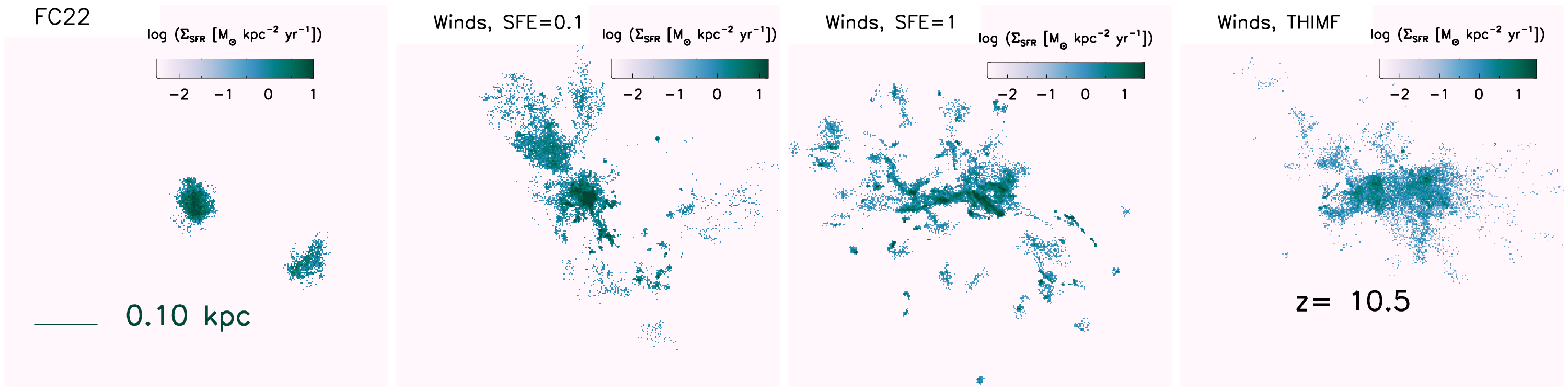}
\caption{Projected SFR surface density maps computed in the x-y plane at $z=10.5$ in the central region of the simulations
  in the FC22 (first panel from left), 'Winds, SFE=0.1' (second), 
  'Winds, SFE=1.0' (third) and 'Winds, THIMF' (fourth) models.} 
\label{fig_map_sfr}
\end{figure*}

\begin{figure}
\center
\includegraphics[width=9.cm,height=9.cm]{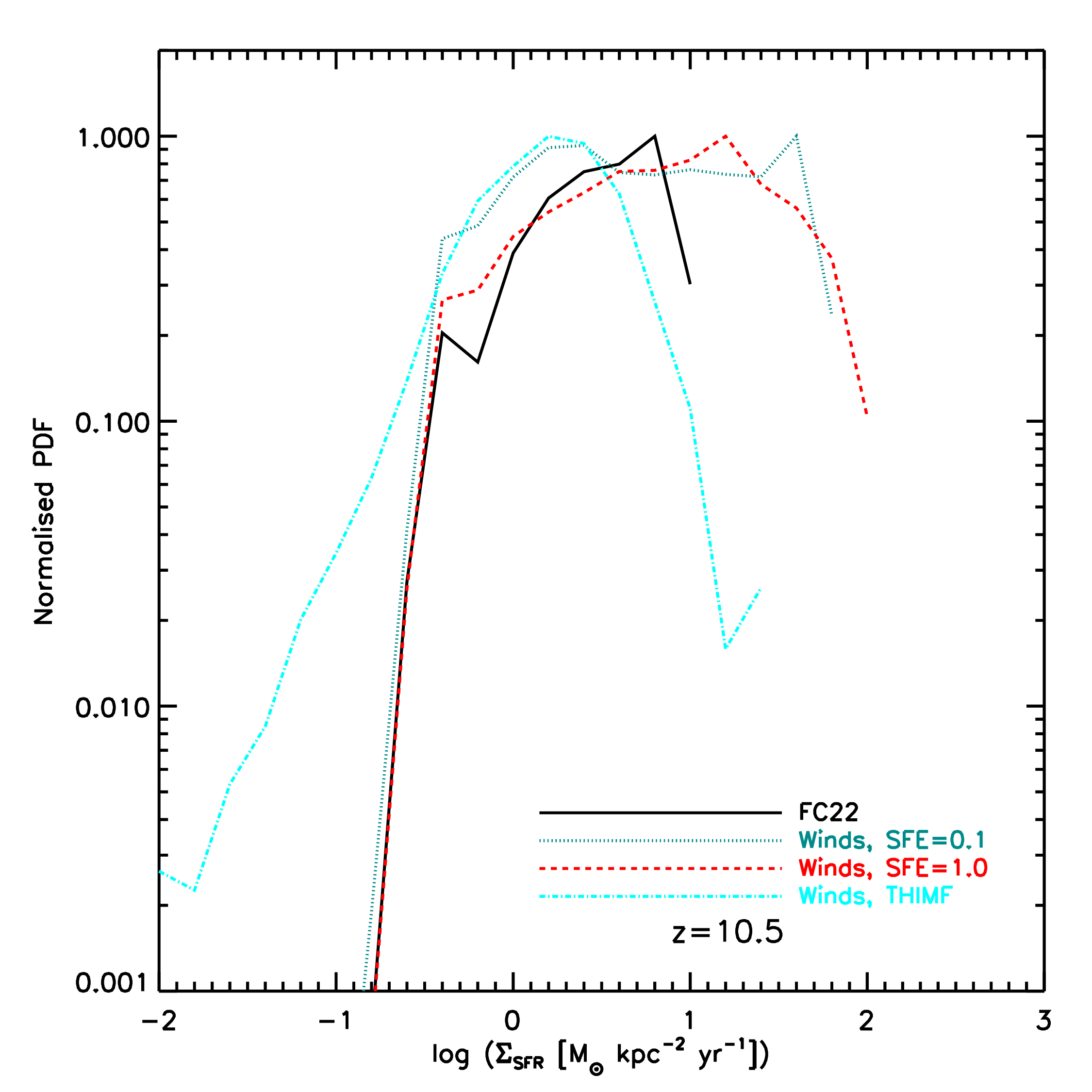}
\caption{Normalised (with respect to the maximum) 
  probability distribution function of the SFR surface density in the same
  central region of our simulations as in Fig. \ref{fig_map_sfr}, at $z=10.5$ and in different
  models.  FC22: black solid line; 'Winds, SFE=0.1':  dotted dark-cyanline;  
  'Winds, SFE=1.0': dashed red line; 'Winds, THIMF': dot-dashed light-cyan line.} 
\label{fig_pdf_SFR}
\end{figure}

\subsection{Properties of the star forming gas}
\label{sec_gas} 
The large-scale (i.e. on 1 -10 kpc-scale) properties of the star-forming gas are not sensitive to the adopted prescriptions regarding
feedback implementation, IMF and  SFE. 
Significant differences in the physical properties of the gas
are instead visible 
at the smallest scales probed by our simulations, as seen in Figures \ref{fig_map_gas}
and \ref{fig_press_gas}, 
showing zoom slice density maps  
and projected gravitational pressure maps at various redshifts, respectively, both
centered on the central clump of each model. 

In the case of the FC22 model, at all redshifts the gas density maps 
(panels in the left column of Fig. \ref{fig_map_gas}) show quite smooth distributions,
in particular in the inner regions of each clump, where the amount of visible overdensities and filaments is
modest. 
The adoption of stellar winds in the pre-SN phase (panels in the second column from left) causes a 
more complicately structured density distribution. 
In particular, at z=15.5, soon after the beginning of SF, the central clump shows a more perturbed distribution than 
the FC22 model, with a density pattern becoming more and more complex as redshift decreases.
This can be appreciated 
from the amount of filamentary structures which populate the central region.  

The new feedback prescriptions adopted here are the key ingredient producing the most remarkable differences with respect to
our previous simulations. 
The maps computed for the SFE=1 and THIMF models 
do not show appreciable differences with respect to the 'Winds, SFE=0.1' model, where
a perturbed pattern including a few knots appears already at $z=13$. 
Another remarkable feature is that, in a few cases,  
each of the three 'Wind' models show maximum densities up to $\sim 10^{-18}$ g cm$^{-3}$, significantly 
larger than the maximum values of the FC22 simulations. 

A clearer, quantitative view of the density structure of our simulations can be seen in Fig.~\ref{fig_pdf_gas},  
where we show the normalised probability distribution function (PDF) of the
gas density of the central region extracted from the density maps of Fig.
\ref{fig_map_gas}, computed for all our models and at different redshifts.
Here, the PDF has been computed from the density in each pixel, and the resulting distribution has been normalised to the maximum value. 
In the FC22 model, at each redshift the density shows a narrow, asymmetric distribution,
peaking at large density values 
($\sim 10^{-20}$ g cm$^{-3}$) and with an extended tail running towards low-density values.
The 'Winds' models (second from top to fourth panel in Fig.~\ref{fig_pdf_gas}) show a broader
  density PDFs and a stronger redshift evolution. 
In particular, the 'Winds, SFE=0.1' model 
is  characterised by a
 significant shift of the peak density  of
nearly two orders of magnitude between $z=13$ and $z=10.5$. 
All cases show a tail on the right of the peak, that extends to larger maximum density values than the FC22 model.   
When compared with each other, the distributions of the 'Winds' models  do not show particularly strong
differences, especially at the lowest redshift,  $z=10.5$, where they peak at similar density values (with a
peak at slightly lower density in the 'Winds, SFE=1' model). 
This remarkable change in shape of the PDFs at various redshifts -from narrow to broad distributions
while transitioning from the FC22 to the 'Winds' models- confirms that the 
stellar feedback prescriptions affect the most the density structure of the star-forming gas,  weakly
affected by other parameters, such as the SFE and the IMF. 

In Fig. \ref{fig_press_gas} we report the gravitational pressure of the gas,
defined as the gravitational force per unit area $P_{\rm grav} = G \times \Sigma_{\rm gas}^2$ \citep{elm97}, 
where $G$ is the gravitational constant and $\Sigma_{\rm gas}$ is the surface density of the gas. 
In general, in virialised, bound systems like dense clumps or star clusters, the gravitational pressure equals
the kinematic pressure, expressed as $\rho~\sigma^2$ 
(where $\sigma$ is the velocity dispersion, \citealt{elm97}, \citealt{ma20}, \citealt{cal22}). 

From Fig. \ref{fig_press_gas} we see that from the very beginning ($z=15.5$),
the gas at the centre of the system is highly pressurized at values 
P$_{\rm grav}$/k  $> 10^8$ K cm$^{-3}$, typical of the  
most turbulent local star-forming regions \citep{sun18,mol20}. 
These pressure values correspond to 
a cloud particle density of $10^4$ cm$^{-3}$ and a turbulent velocity of $10$ km/s. For comparison, such values
are several orders of magnitudes 
larger than those of the diffuse, warm ISM in the Milky Way disc, which has typically P$_{\rm grav}$/k  $\sim 10^3$ K cm$^{-3}$.

One significant difference between the FC22 and the 'Winds, SFE=0.1' model is that in the latter, the high-pressure
(P$_{\rm grav}$/k$>10^8$ K cm$^{-3}$) regions are significantly more compact and showing a complex, filamentary
pattern, consisting of multiple small-size knots, at variance with the smoother distribution of the former.
It is worth noting that in the 'Winds, SFE=0.1', the high pressure-gas is distributed more 
widely than in the FC22 model. 
At z=10.5 the high pressure-gas distribution reflects the young stars (Fig.~\ref{fig_map_sfr}) and total stellar distribution (Fig. \ref{fig_map_stars}). 
Essentially, more stellar aggregates were born in the 'Winds, SFE=0.1' model with respect to the FC22 one, which are more scattered
from their birth and that can also undergo significant dynamical interaction, and within which the youngest and most massive stars affect
the surrounding gas with their feedback. 

At later epochs, the FC22 and the three 'Winds' models present maximum values up to P$_{\rm grav}$/k  $\gtrsim 10^{9}$ K cm$^{-3}$.

In figure \ref{fig_press_gas}, we also report the 1D density-weighted velocity dispersion of the cold medium
  (with temperature $<200$ K), calculated at $z=10.5$. This quantity is a measure of the turbulence of the gas and can be defined as:  
\begin{equation}
  \sigma_{\rm 1D}^2 = \frac{1}{3}\frac{\Sigma \rho_c [(v_x - \bar{v_x})^2 + (v_y - \bar{v_y})^2 + (v_z - \bar{v_z})^2]}{\Sigma \rho_c}
  \label{eq_sigma}
\end{equation}
(e. g., \citealt{she10}; \citealt{cal20}). In Eq. \ref{eq_sigma}, $\rho_c$, $v_x$, $v_y$ and $v_z$ are the density, 
the x-, y- and z-component of the velocity of the cold  gas 
in a cell respectively, 
whereas $\bar{v_x}$, $\bar{v_y}$ and $\bar{v_z}$ are the average x-, y- and z-component velocity values, respectively. 
Our results show a factor $\sim$2 difference 
in $\sigma_{\rm 1D}$ between the FC22 and Winds models, a feature that can be ascribed to the stronger feedback that characterises the former. 

Moreover, the computed values are very similar in all the 'Winds' models, indicating that this quantity
is not sensitive to the SFE and IMF, and are of the order of  
the velocity dispersion measured in local clouds, e. g. as traced by the relation between $\sigma$ and size observed in 
molecular clouds (\citealt{lar81}, \citealt{elm00}). The computed values are larger than the thermal velocity dispersion 
of the cold gas, which is of the order of $1$ km/s and corresponding to the sound speed at the considered minimum temperatures ($\sim 100 $ K).
This is the signature of a significant degree of turbulence, that does not change significantly with the different prescriptions
adopted in our study. 

The fundamental reason for the smoother density and pressure maps obtained with the FC22 prescriptions  
is the stronger, prompt stellar feedback that characterizes this model. In this case, 
the initial, instantaneous injection of thermal energy coincident with massive star formation 
wipes away all the small-scale structures in both the density and pressure fields,
smoothing out the spatial distributions and creating the homogeneous patterns visible in Figs. \ref{fig_map_gas} and \ref{fig_press_gas}.

The properties of the central star-forming regions are illustrated further by the star formation rate (SFR) density
maps of Fig.~\ref{fig_map_sfr}, that represents a projected map computed considering the youngest stars at $z=10.5$.
In each snapshot, we have considerered the stars younger than $15$ Myr\footnote{High mass stars younger
  than $\sim 10-15$ Myr are
  responsible for powering strong emission lines such as H$\alpha$ and Lyman $\alpha$, frequently used as star formation indicators 
  (e. g., \citealt{vil15,oya16}).}
and, in each pixel, computed their cumulative 
mass and divided it by this time interval and the pixel area.
The distribution of young stars in the SFR maps reflects the morphology of the
highest pressure regions in Fig. \ref{fig_press_gas}.
In all cases, the highest SFR values are visible at the centre of the clumps, where the youngest stars
can be found.\\
 Additional information on SFR variations in different models is provided in Fig. \ref{fig_pdf_SFR}, showing 
the normalised 
PDFs of the SFR density in our models, all computed from the $\Sigma_{\rm SFR}$ values in each pixel at $z=10.5$. 
The FC22 and Winds, SFE=0.1/1.0 models show remarkable similarities at the lowest SFR values, $\dot{\Sigma}_{\rm *} \le 1$ M$_{\odot}$ yr$^{-1}$, 
kpc$^{-2}$ and marked differences at larger values.
The FC22 model shows a sharp peak at $\dot{\Sigma}_{\rm *} \sim 6-8$ M$_{\odot}$ yr$^{-1}$ kpc$^{-2}$, whereas the 
'Winds' models computed with a \cite{kro01} IMF show more flat-topped distributions. 
The 'Winds, SFE=1.0' shows a broad peak at $\dot{\Sigma}_{\rm *} \sim 10$ M$_{\odot}$ yr$^{-1}$ kpc$^{-2}$ and extends
up to larger values (up to $\dot{\Sigma}_{\rm *} > 50$ M$_{\odot}$ yr$^{-1}$ kpc$^{-2}$) than its lower-efficiency homologous.
On the other hand, a striking feature of the 'Winds, THIMF' model distribution is that it peaks at
a lower $\dot{\Sigma}_{\rm *}$ 
value ($\dot{\Sigma}_{\rm *} \sim 1$ M$_{\odot}$ yr$^{-1}$ kpc$^{-2}$) than the FC22. Moreover, the lowest SFR bins
($<0.3$ M$_{\odot}$ yr$^{-1}$ kpc$^{-2}$) 
are more populated than in all the other cases, underscored by the fact that its normalised distribution shows
the highest values in this regime. 
The star formation history (SFH) of the central region in all our models is illustrated in Fig. \ref{fig_sfh},
along with the evolution of the cumulative stellar mass obtained in the different cases. 
This figure is helpful to highlight the regulating effects of stellar feedback on the SFH in the different models. 

\begin{figure*}
\center
\includegraphics[width=18.cm,height=4.5cm]{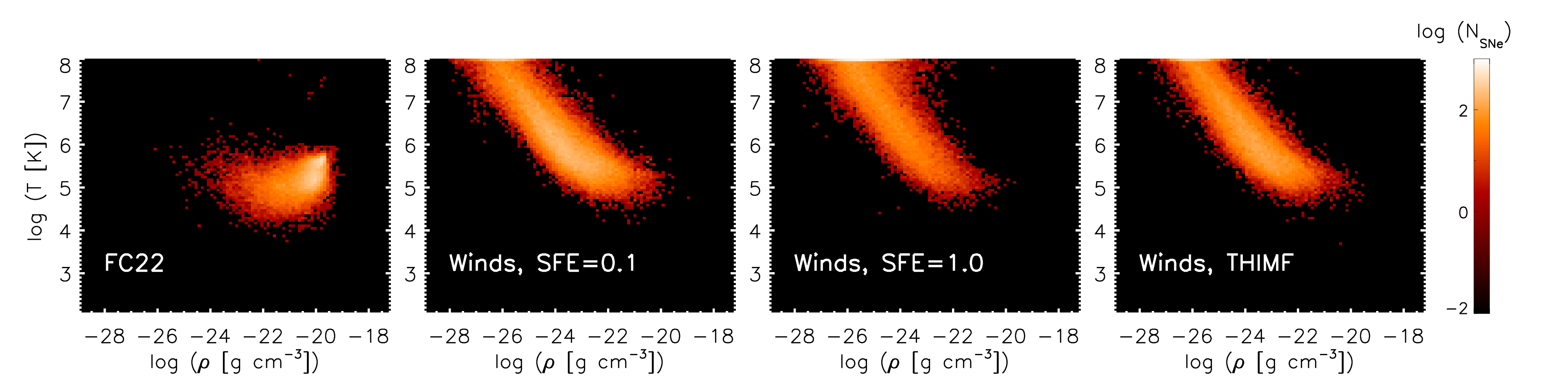}
\caption{Temperature-density diagrams of the gas in which SNe explode in different models (see the text
  for further details) in the FC22 (first panel from left), 'Winds, SFE=0.1' (second), 
  'Winds, SFE=1.0' (third) and 'Winds, THIMF' (fourth) models. The colour scale correspods to the number of SNe exploding
  in each pixel of the map.} 
\label{fig_SNphase}
\end{figure*}
\begin{figure*}
\center
\includegraphics[width=18.cm,height=4.5cm]{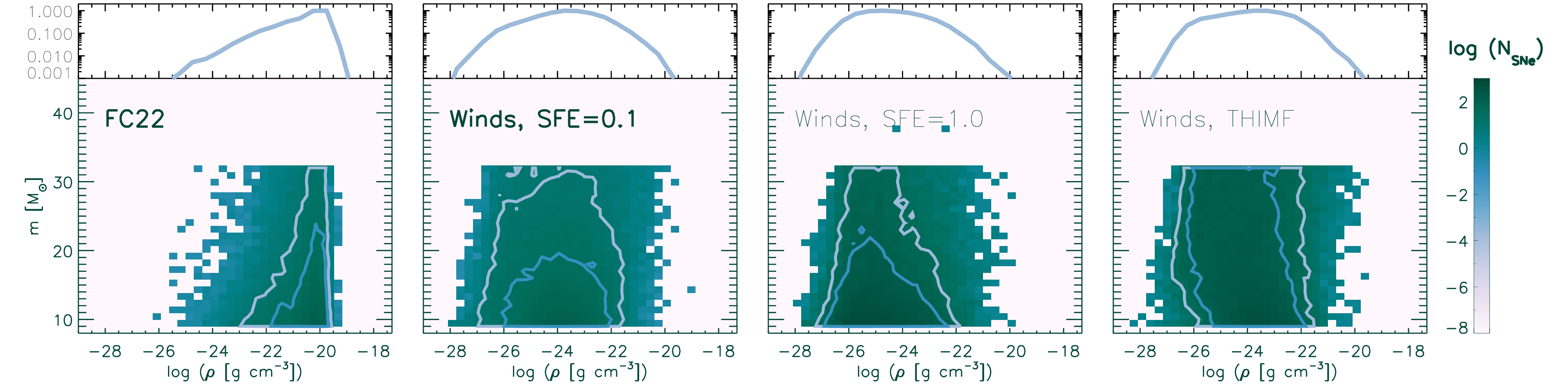}
\caption{Density of the gas in which SN explode vs SN progenitor mass in the FC22
  (first panel from left), 'Winds, SFE=0.1' (second), 
  'Winds, SFE=1.0' (third) and 'Winds, THIMF' (fourth) models. 
  The colour scale illustrates the number of SNe exploding 
  in each pixel of the map.  The light-blue and blue thick contours enclose numbers of
  SNe corresponding to 60\% and 75\% of the maximum, respectively. 
  The histograms on top of each panel are the normalised PDFs of the density. 
} 
\label{fig_SNrho}
\end{figure*}
\begin{figure*}
\center
\includegraphics[width=18.cm,height=4.5cm]{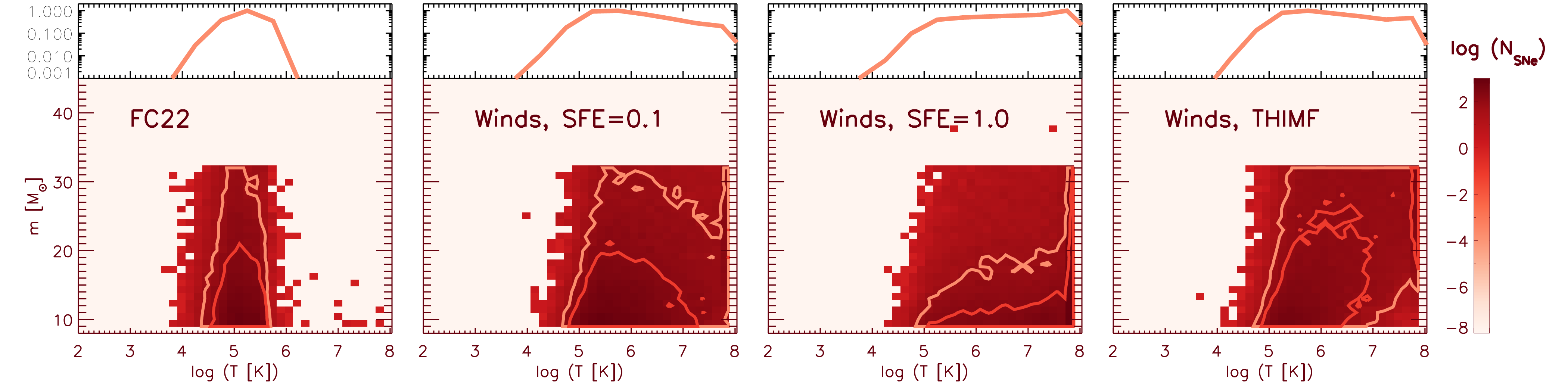}
\caption{Temperature of the gas in which SN explode vs SN progenitor mass in the FC22
  (first panel from left), 'Winds, SFE=0.1' (second), 
  'Winds, SFE=1.0' (third) and 'Winds, THIMF' (fourth) models.
  The colour scale illustrates the number of SNe exploding
  in each pixel of the map. 
  The light-orange and red thick contours enclose numbers of
  SNe corresponding to 60\% and 75\% of the maximum, respectively.
  The histograms on top of each panel are the normalised PDFs of the temperature. } 
\label{fig_SNT}
\end{figure*}

\subsection{Properties of the Pre-SN feedback-driven gas}

In their pre-SNe phase, massive stars play a fundamental role in determining the physical state of the gas
and of the properties of the stellar aggregates.  The pre-SN feedback 
affects the properties of the gas immediately after the formation of new massive stars.
On the other hand, SN explosions occur at the end of the stellar lifetimes,
therefore SNe of different masses may find diverse gas conditions. 
In general, SNe of higher masses explode soon after the formation of a new stellar generation. When they explode, 
the properties of the gas may be closer to the ones of the star-forming gas. 
On the other hand, lower-mass SNe explode later, when the conditions of the gas may be in principle  
affected by previous explosions. 
SNe have little effects on the evolution and star formation history of the single clusters.
  This will be shown in forthcoming works that will address the SFHs of the clusters
  (Pascale et al. 2025; Ragagnin et al., in prep.). 

To investigate further how our prescriptions affect the gas,
we study the differences in the physical properties of the medium where SNe explode in various models. 

In Fig. ~\ref{fig_SNphase}, we show 2-dimensional temperature-density distributions
of the gas where SNe of various masses have exploded in our four different models\footnote{The density and the temperature values shown
  in Fig. ~\ref{fig_SNphase} are the ones of the gas in the cell, one timestep before the explosion affects the ISM.}.  
The FC22 model shows a narrow distribution, indicating rather similar conditions in which SN explode,
with a $\sim 2$ and $4$ orders of magnitude dispersion for T and $\rho$, respectively. 
On the other hand, the 'Winds, SFE=0.1' model shows a clear anticorrelation that is more similar
to typical phase-diagrams of a feedback-driven ISM \citep{ros17,fel23,gur24}. 
The 'Winds, SFE=0.1', 'Winds, SFE=1.0' and 'Winds, THIMF' models show very similar temperature-density distributions,
supporting that the change in feedback prescriptions, from the explosive early feedback of FC22 to the more gentle, stellar
winds adopted here, plays a stronger role in driving the relation of Fig. ~\ref{fig_SNphase} than the other parameters. \\
As already shown in FC22 (see the phase diagrams in Fig. 4 of the supplementary material of \citealt{cal22}),
 the cold, dense gas with T $<2 \times 10^4$ K is promptly turned into new stars. 
 This is the reason why very few SNe explode below this temperature, together with the fact that in the 'Winds' models,
 massive stars heat continuously the gas before the end of their lifetimes.
 \\ 
Further information on the SN-driven gas is provided in Fig.~\ref{fig_SNrho}, illustrating 
the density of the gas in which SNe of decreasing mass explode. 
In the FC22 model, SNe explode mostly in relatively dense gas, with the computed distribution peaking 
slightly below $\rho \sim 10^{-20}$ g cm$^{-3}$. This is the effect of the long time elapsed between
early pre-SN feedback, occurred instantaneously after star formation, and SN explosions, during which
the gas is essentially allowed to re-condense and cool. 
As highlighted by the light-green and blue thick contours, enclosing numbers of SNe corresponding to 60\% and 75\%
of the maximum, respectively, the density range broadens towards lower densities with decreasing mass. 
This means that, at later times, supernovae can explode under a wider range of physical conditions. 
This is confirmed also by the left panel of Fig.~\ref{fig_SNT}, 
showing the relation between SN mass and temperature at the explosion, 
showing a relatively narrow T distribution, peaking at T$\sim 10^5$ K and broadening towards lower masses. 
The 'Winds' models show significantly different distributions in the SN mass-density plane
and density PDF as well (shown on the top of each plot in Fig.~\ref{fig_SNrho}). The 'Winds, SFE=0.1' model shows
a broad, more symmetric density distribution, with SNe exploding at peak 
 density $\rho \sim 10^{-24}$ g cm$^{-3}$, with an overall weaker correlation between mass and density.
 In the same model, the T distribution of Fig.~\ref{fig_SNT} is strongly asymmetric, due to
 the fact that SNe cluster at T=10$^8$ K, the value chosen for the maximum gas temperature.
 The saturation at T=10$^8$ K  also indicates that in this model, the star formation is significantly more concentrated
 than in the FC22 one. This effect is even stronger for the 'Winds, SFE=1.0' model,
 in which the SN explosion density peaks at about $\rho \sim 10^{-26}$ g cm$^{-3}$, i. e. at significantly
 lower density, whereas the T distribution shows a stronger clustering at T=10$^8$ K.
 This aspect highlights that also the SF efficiency plays an important role in shaping the $\rho$ and T structure
 of the feedback-driven gas.  
 Finally, the 'Winds, THIMF' is the model that shows the clearest sign of a correlation between exploding SN mass and T,
 highlighted by the two thick contours in the rightmost panel of Fig.~\ref{fig_SNT}. 
 This is the result of the strongest early feedback achieved in this case with respect to the other 'Winds' models,
 in which the final cumulative  stellar mass is the lowest (see Tab. ~\ref{table_sim} and Fig.~\ref{fig_sfh}), the 
 stars are the least concentrated and where, 
 when the lowest mass SNe explode, the early effects of 
 pre-SNe have weakened the most (although the hint for a similar behaviour is shown also by the 'Winds, SFE=0.1' model).
 
\begin{figure*}
\center
\includegraphics[width=17.cm,height=12.75cm]{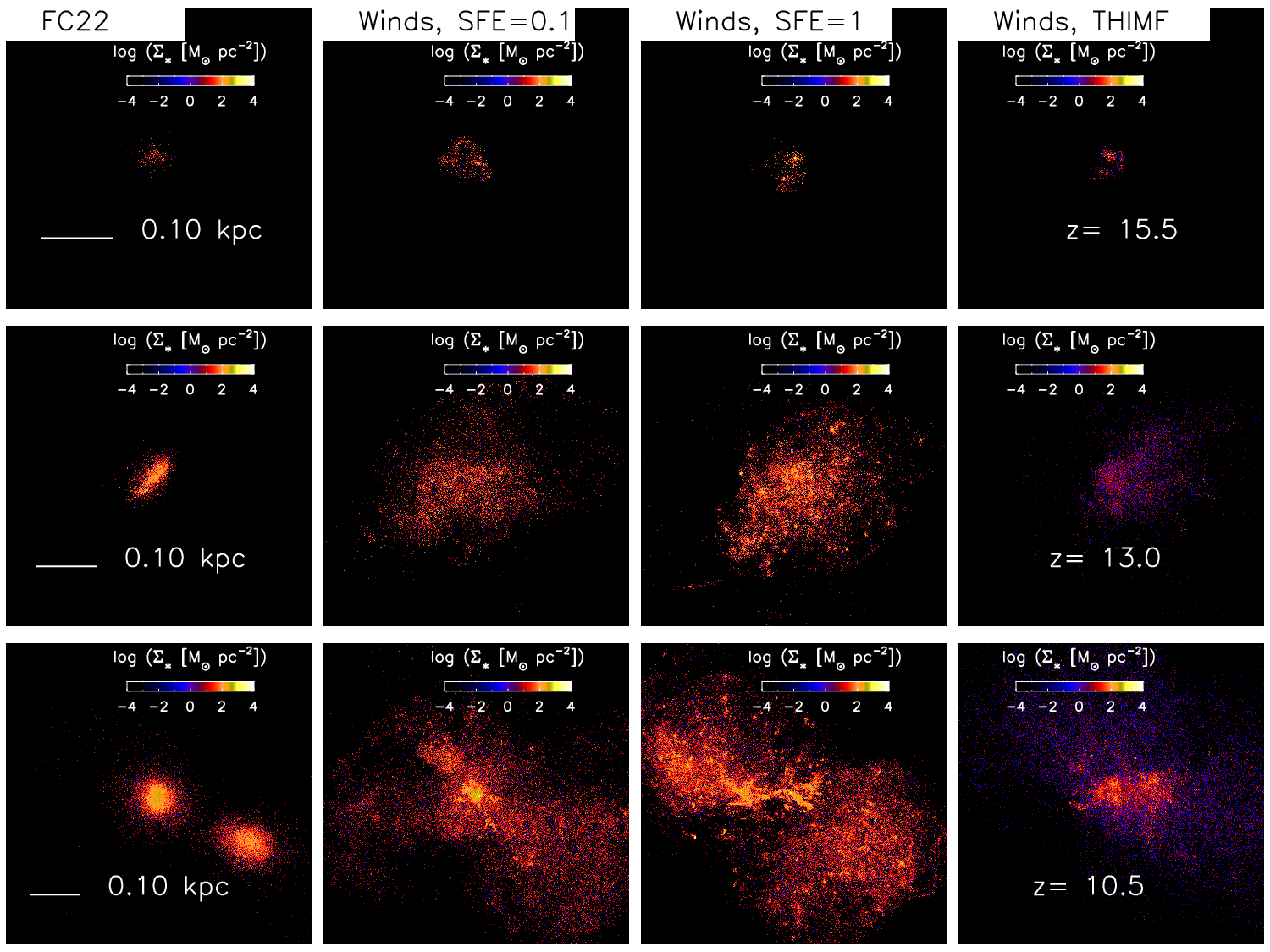}
\caption{Projected stellar density map in the x-y plane in the four models presented in Table \ref{table_sim} 
  at different redshifts. 
  The maps describe the stellar density in the central region of the simulations in the 
  FC22 (first column from left), 'Winds, SFE=0.1' (second), 
  'Winds, SFE=1.0' (third) and 'Winds, THIMF' models (fourth column).  
  The maps are at $z=15.5$ (upper row), $z=13$ (middle row) and  $z=10.5$ (bottom row). 
  The horizontal white solid lines shown in the left-hand column-panels indicate the physical scale.} 
\label{fig_map_stars}
\end{figure*}

\subsection{Properties of the stellar clumps}

The stellar mass density maps of the central region of our simulations at different redshifts is shown in 
Fig. ~\ref{fig_map_stars}.
Starting from the earliest time ($z=15.5)$, the FC22 model shows a diffuse central
clump with a maximum density of $10^2 M_{\odot}$ pc$^{-2}$. At later epochs,
the mass and size of the central clumps grows significantly, but the same is not true for the maximum
stellar density. At $z=10.5$ the stellar clumps show extended and smooth distributions, reflecting
the properties of the gas discussed in Sect. \ref{sec_gas}, and with overall sizes of approximately $\sim 100$ pc.
At the earliest redshift, the 'Winds, SFE=0.1' model does not show significant differences with the FC22,
barring a more conspicuous amount of stars, becoming more noticeable at $z=13$. 
At  $z=10.5$, the 'Winds, SFE=0.1' shows more a extended stellar distribution than FC22, with a few dense clumps and a
significant amount of diffuse stars. The maximum density values achieved in this model are higher
than the FC22 by more than 1 order of magnitude.  

On the other hand, the 'Winds, SFE=1.0' model shows remarkably high stellar density values already at early times,
with two compact stellar clumps clearly visible at $z=15.5$. 
At lower redshift, this model shows some similarities with its lower-SFE homologous,
in particular in terms of the extent of the diffuse stellar component but, besides the larger $\Sigma_{\rm *}$ values,
it also shows a larger abundance of dense aggregates at $z=10.5$.

The stellar component in the 'Winds, THIMF' model shows intermediate features between the 'Winds' and the 'FC22' models,
with a distribution of diffuse stars similar to the one of the 'Winds, SFE=0.1', but with the presence of very few
compact knots with maximum densities similar to their counterparts in the FC22 model.

The 'Winds-THIMF' model is characterised by a much stronger feedback (i.e. a higher injected specific mass and energy rate, see Fig. 1)
  than the 'Winds, SFE=0.1' and the 'Winds, SFE=1.0' models, characterised by 
  a standard \cite{kro01} IMF. This is the reason for the significantly lower density of the star clusters present in this model. 

A few more representative star clusters and clumps in our 4 models are shown in Fig. \ref{fig_map_clumps}
and will be discussed in App.  \ref{sec_more_clumps}. 

\begin{figure}
\includegraphics[width=9.5cm,height=9.5cm]{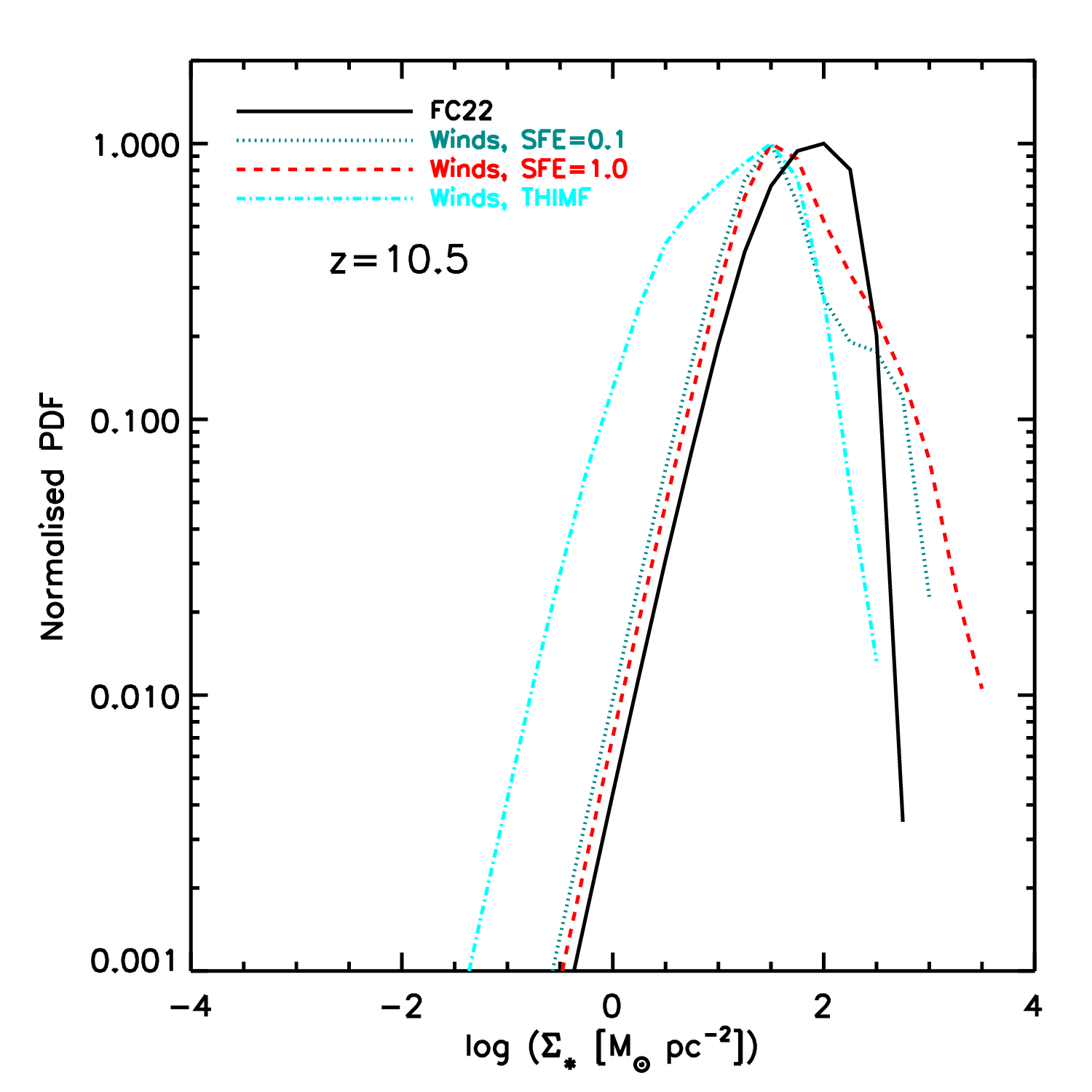}

\caption{Normalised (with respect to the maximum) 
  probability distribution function of the stellar surface density in the same
  central region of our simulations as in Fig. \ref{fig_map_sfr}, at $z=10.5$ and in different
  models.  FC22: black solid line; 'Winds, SFE=0.1': dotted dark-cyan line;  
  'Winds, SFE=1.0': dashed red line; 'Winds, THIMF': dot-dashed light-cyan line.} 
\label{fig_pdf_sigma}
\end{figure}

\begin{figure}
\includegraphics[width=9.5cm,height=7.12cm]{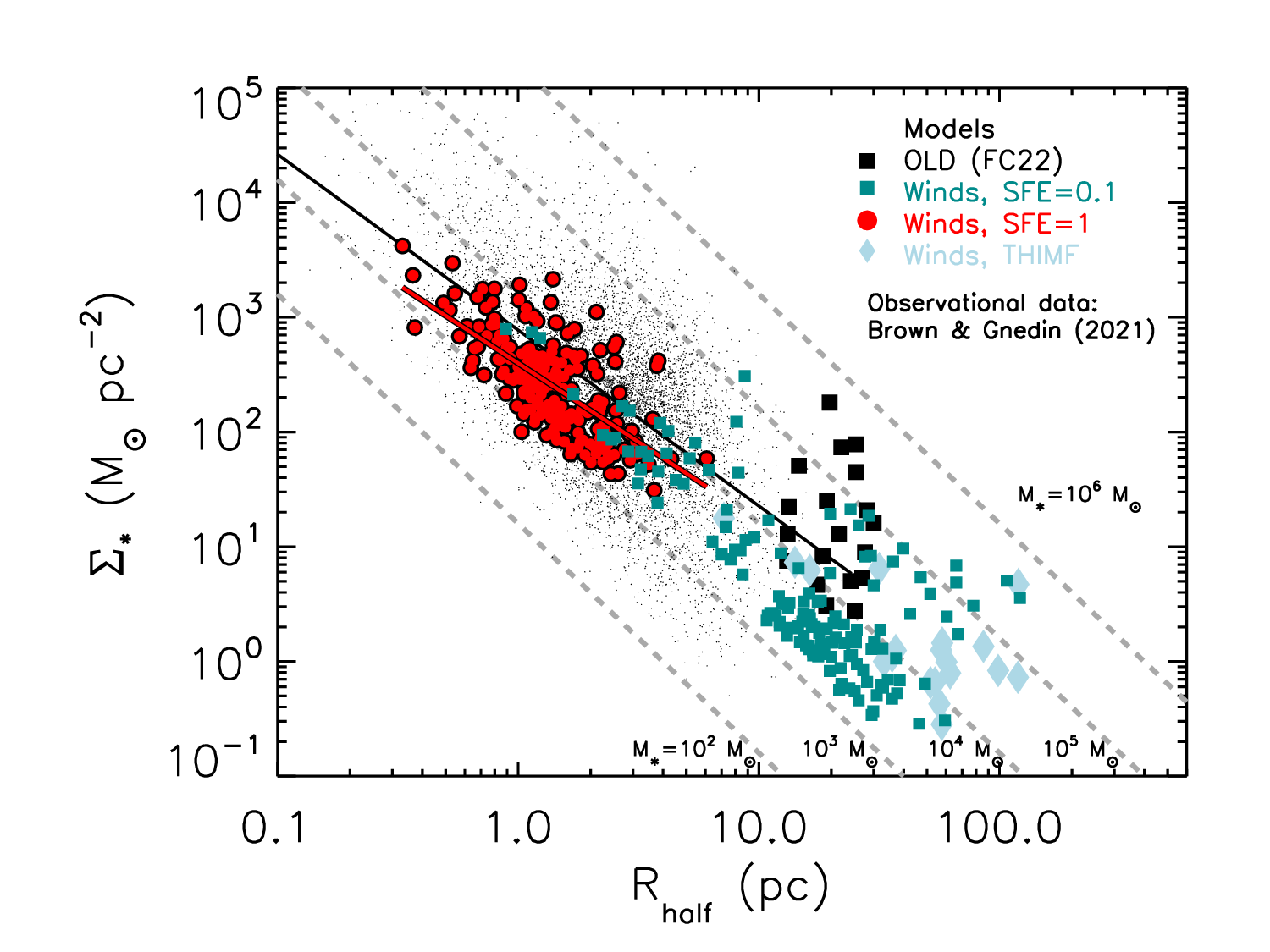}
\caption{Relation between stellar surface density and size (defined as half-mass or half-light radius)
  in our simulations and as observed in local star clusters. The black solid squares, dark-cyan solid squares,
  red solid circles and light-cyan diamonds are the star clusters or clumps retrieved at $z=10.5$ in the FC22, 'Winds, SFE=0.1', 'Winds, SFE=1.0'
  and 'Winds, THIMF' model, respectively, and the small black dots are the observational dataset of young star clusters from the Legacy Extragalactic
  UV Survey \citep{ryo17,bro21}. The thick red and thin black line is a linear fit (in log($\Sigma_*$) - log(R$_{\rm half}$) space) to the relation found in the 'Winds, SFE=1.0' model and in the
  observational dataset, respectively. Each grey dashed line represents the relation between $\Sigma_*$ and R$_{\rm half}$
  at fixed stellar mass, for which we have considered various values 
  between $10^2$ M$_\odot$ and $10^6$ M$_\odot$.  } 
\label{fig_sigma_reff}
\end{figure}

The normalised PDF of the central surface stellar density at $z=10.5$ shown in Fig. \ref{fig_pdf_sigma} (computed as in Fig. \ref{fig_pdf_SFR}), 
helps gaining further insight into the differences between 
our models. This plot confirms the narrow distribution of the 'FC22' model, peaking slightly below 
$\Sigma_{\rm *}\sim 10^2~M_{\odot}$ pc$^{-2}$. 
The 'Winds, SFE=0.1' peaks at a value of  $\Sigma_{\rm *}\sim 20~M_{\odot}$ pc$^{-2}$ but is slighly more populated
at the highest values than the 'FC22'. 
The 'Winds, THIMF' and the 'Winds, SFE=1.0' have distribtions skewed towards the lowest and highest
$\Sigma_{\rm *}$ values, respectively. 
These results confirm that the 'Winds, SFE=1.0' model shows the densest stellar aggregates.

\subsection{Comparison with the observed young star clusters}
In  Fig. \ref{fig_sigma_reff} we analyse a standard scaling relation 
for young stellar aggregates, the one between stellar surface density and size. 
In the simulations, the clusters have been identified by means of the Hierarchical Density-Based Spatial Clustering of
Applications with Noise (HDBSCAN) software library  \citep{mci17}.
HDBSCAN builds a minimum spanning tree (MST) based on the distances of the data points, then it uses a
density threshold to build a hierarchy of clusters, set to $25~$M$_{\odot}$ pc$^{-3}$.
This value was chosen on an empirical basis, as we verified that lower values tend 
to classify non-genuine agglomerates as clusters, while higher values lead to the exclusion of the densest structures.
HDBSCAN also requires the expected number of objects per cluster, in our case set to 500, having checked the robustness 
of the results by varying this parameter across a reasonable range. 

For the clustering algorithms, the most common problem is the clump identification 
in regions that contain diffuse, unbound  stellar components. This makes the clump-finding process prone to noise and
identification of unreal systems, such as diffuse, extended regions regarded by the software as clumps. 
The cleaning of the sample from such effects is often troublesome, with authomatic procedures not capable of offering reliable solutions. 
To tackle this aspect, we performed visual inspection of the clumps identified by the algorithm,
ensuring that the identified set of clumps is a reliable one and that the 
rejected clumps were effectively unphysical or classifiable as too diffuse systems.

As for the sizes, in the case of our systems we consider the half-mass radii, computed from the 2D mass density profiles and, for each,
calculating the average values along different projections as performed in FC22. 
In Fig. \ref{fig_sigma_reff} we show the stellar density ($\Sigma_{\rm *}$\footnote{We define the surface stellar density
  $\Sigma_{\rm *}=\frac{M_{\rm *}}{2 \pi R_{\rm half}^2}$, where $M_{\rm *}$ and $R_{\rm half}$ are the cluster stellar mass and half-mass radius, respectively.})-size relation for the samples of clusters and clumps identified
at z=10.5 in all our models. 
Although the properties of the star-forming ISM are known to depend critically on redshift \citep{tac20}, 
without any observational knowledge of the evolution of the size and mass of young stars clusters across cosmic time,
we compare the model properties with an observational dataset of young star clusters from the Legacy Extragalactic UV Survey \citep{ryo17}
collected by \cite{bro21}. 
The latter is essentially the largest database of local young star clusters (YSCs) radii currently available.  
In this dataset, the sizes indicated with $R_{\rm half}$ are the half-light radii.

The clumps obtained in the 'FC22' model (black solid squares) are very diffuse and 
incompatible with local YSCs. 
The systems identified in the 'Winds, SFE=0.1' model (dark-cyan squares) build a diagonal
sequence that is partially in agreement with the local sample, yet with too few clusters reaching a density high enough 
to account for the observational relation of  YSCs,  populating the upper-left area of the diagram.
Moreover, a significant fraction of the clumps identified in the  'Winds, SFE=0.1' extend to size values $>10$ pc and 
have densities $<10$ M$_{\odot}$ pc$^{-2}$, 
without any observational counterpart.  
On the other hand, the 'Winds, SFE=1.0' model allows us to reproduce nicely the local YSCs relation, as underscored further by the
agreement of the linear fits (in log-scales) obtained for the observational data and simulated clusters, represented by the black and red solid lines,
respectively. 

A comparison between the results of the 'Winds, SFE=0.1' (dark cyan squares) and FC22 model (black squares) shows that,
for a given IMF, the release of feedback in the form of stellar winds produces significantly denser structures. 
A comparison between the results of the 'Winds, SFE=1.0' (red circles) and the 'Winds, SFE=0.1' (dark cyan squares)
shows that, for the same feedback implementation and IMF, the adoption of a higher SFE produces denser star clusters. 
The shape of the observational distribution is accounted for, 
with our 'Winds, SFE=1.0' model showing the presence of a few systems with average densities up to
several $10^3$ M$_{\odot}$ pc$^{-2}$. 
Still, the less frequent, observed average densities of $\sim 10^4 - 10^5$ M$_{\odot}$ pc$^{-2}$ shown by some YSCs are
not reproduced, 
and this requires further investigation in future works. 
Finally, we note that the THIMF model produces the most diffuse clumps, even more diffuse that those 
obtained in FC22. 
The result of extremely loose star clusters with a THIMF is the very strong pre-SN feedback characterising this model
  (Fig.~\ref{fig_stellar}), in which the specific cumulative mass and energy for a THIMF (red dotted lines in Fig. 1) are
  significantly higher than with a K01 IMF, to become  
  comparable to the ones of the 'Winds, FC22' model at $\sim$ 5-6 Myr and the strongest of all models at later times.  
This is also the reason of the similarity of the cumulative star formation history of the C22 and THIMF model. 

To summarise the results of this Section, the adoption of a high SFE is the key ingredient that allows us to
eventually achieve realistic star clusters in our simulations, with properties similar to the ones of local YSCs.

\section{Discussion}
We have shown how both stellar feedback and star formation efficiency play a fundamental
role to obtain high-density systems that can be classified as star clusters.
In this Section, we discuss in more detail these fundamental aspects and the implication of our
results in a more general framework, also comparing them with those obtained in other studies.

\subsection{Importance of the Pre-SN feedback}
Recent studies have shown that early feedback from winds and radiation, occurring
before supernova explosions, is necessary to account for some fundamental properties of 
GMCs,  such as the 'de-correlation' between molecular gas and young stellar regions at $\sim$100 pc scales \citep{che20}. 
The fact that the co-existence of GMCs and H II regions is very rare on such scales indicates that the 
evolutionary cycling between GMCs, star formation, and feedback must be rapid and efficient \citep{kru19},
and that early feedback is the dominant 
process that drives the destruction of molecular clouds, on typical timescales of a few Myr \citep{che20},
before the first SN explosions occur. 

In the process of pre-SN feedback, the energetic input from massive stars is known to  
play a crucial role in pre-processing the gas before SNe explode. 
In fact, while SN explosions are known to dominate the total energy budget on timescales comparable to the lives of massive stars, i.e.
up to $\sim 20-30$ Myr (e. g., \citealt{fic22}, see also Fig. 1), the pre-SN feedback is of great importance for reducing 
the circumstellar gas density and limiting radiative losses in SN remnants, strenghtening their impact.

Traditionally, in cosmological simulations pre-SN feedback has often been ignored 
since it acts on scales much smaller than the typical maximum resolution 
(typically $>50$ pc) that can be achieved in a fully cosmological framework,
although a few previous studies have recognised its role, modelled with crude, sub-grid approximations \citep{hop11,sti13,age13}.

Various pre-SN feedback processes are known to act on scales betwen $\sim 1$ pc and $\sim 10$ pc (\citealt{fic22} and references therein). 
These scales can be resolved mostly in non-cosmological simulations, which typically represent isolated 
systems such as star clusters (e. g., \citealt{cal15,yag22}) or dwarf galaxies \citep{age20,lah23}, where the role of such ingredients can be investigated 
more carefully. 
Our results highlight further the importance of pre-SN stellar feedback  as a main 
driver of the early evolution of star clusters. 
In our previous work, we modelled pre-SN feedback as a local, impulsive release
of thermal energy and mass from each newly formed massive star, likely representing an overestimate of its effects and leading
to excessively diffuse stellar clumps. 
In our new simulations, we have shown that the prescription of stellar winds in the form of continuous,
slow injection of thermal energy and mass plays a fundamental role in increasing the compactness of the
stellar aggregates.

In simulations, the development of the wind-blown bubbles depends crucially on numerical resolution (e. g., \citealt{lan21}). 
\cite{pit21} discussed the required criteria to inflate a stellar wind bubble in different conditions,
such as via momentum and thermal energy injections. 
In the case of pure thermal energy as in our case, the pressure in the
injection region should exceed the environmental one $P_{\rm amb}$; this criterion corresponds to a
well-defined requirement for the maximum size of the injection region as a function of the mass return rate $\dot{m}$ and the terminal velocity of the wind
$v_w$: 
\begin{equation}
r_{\rm inj} = \left( \frac{\dot{m} v_w}{4 \pi P_{\rm amb}} \right)^{1/2}. 
\end{equation}
In our case, considering the very high gas density values of $>10^5$ cm$^{-3}$ of our star-forming regions and $P_{\rm amb} \sim \rho~T~ k_{\rm B}/m_p$,
a 0.02 pc cell width is required to resolve the wind-driven bubbles, i.e. a factor 10 smaller than our actual resolution at $z\sim 16$, 
therefore we have to recur to our sub-grid, 'delayed-cooling' implementation. 
In other simulations aimed at modelling star cluster formation, 
stellar feedback is often implemented in the form of momentum injection \citep{kim16,li19}.  
Despite it has been shown that the clustering of the feedback sources enhances their effects in terms of both momentum and energy
\citep{yad17,sch18}, in most cases for the feedback to be efficient this choice requires the momentum to be artificially boosted by a given factor, 
that can reach values in excess of 100 even at our resolution \citep{pit21}. \\
However, we are aware that care needs to be taken when treating stellar winds as the primary feedback process on
the typical GMCs scale. 
In turbulent clouds, most of the energy deposited by stellar winds is dissipated by various processes, including 
efficient radiative cooling through the mixing between the hot, shocked medium and the colder, surrounding gas
\citep{lan21}. 
In such systems, ionising radiation from young massive stars is more effective
at regulating  SF in the pre-SN phase  \citep{wal12,ros15,gee15,kim18}. 
Moreover, also in this case metallicity is expected to have a relevant effect and, 
according to previous studies of isolated turbulent clouds, it may affect the SFE as well \citep{he19}. \\
Although the exact details have not been discussed explicitly, in cosmological simulations 
this ingredient seems to make a difference even in presence of momentum injection
from stellar winds \citep{kim16,ma20}.

In dense environments, also radiation pressure may play an important role in regulating the early phases of star formation
in young massive clusters \citep{fal10}, even though its regulating effect on GMC scales has been questioned
(e. g., \citealt{meno22}). 
 
In the present framework, the contribution of ionising radiation and radiation pressure from
individual massive stars remains to be tested, to check how it impacts the star cluster density and whether it requires
further artificial assumptions to work, such as momentum boosting;  
this will be the focus of a future work of the SIEGE series.

\subsection{A high star formation efficiency in dense stellar systems}
The second important ingredient to achieve dense star clusters is a high SFE. 
The results of our study  
indicate that a high SFE, defined as the fraction of gas converted into stars per
freefall time $\epsilon_{\rm ff}$  (see Eq.  2),  is needed to reproduce the density values observed in local young
star clusters. 
Considering a density of $10^5$ particles cm$^{-3}$, typical of our star-forming gas, $\epsilon_{\rm ff}=1$
  corresponds to very fast SF timescales of $\sim~0.2$ Myr. 
In the present framework, it is important to stress that the SFE is an assumed quantity.
The capability to predict a possible range of values for this quantity with an ab-initio approach
requires a different type of simulations, including more physical processes, such as magnetohydrodynamics, and
spatial  resolution down to the AU, currently unfeasible in our conditions, but marginally achievable
in studies of isolated SF regions (e. g. \citealt{gru19,pol24}). \\

It is worth noting that other definitions of SFE can be found in the literature.
A popular one is $\epsilon$, sometimes referred to as the instantaneous SFE 
\begin{equation}
 \epsilon = \frac{M_{\rm *}}{M_{\rm *} + M_{\rm gas}},  
\end{equation}
where $M_{\rm *}$ is the stellar mass and $M_{\rm gas}$ is the gas mass associated
with the star-forming cloud, 
often represented by the molecular gas and inferred
  via suitable tracers \citep{gru19}. 

Observations suggest that in local star-forming regions, including dense clumps and giant molecular clouds,
the SFE can show significant scatter, from a few tenths of percent to a few percent (see Tab. 2 of  \citealt{gru19}).
It is worth noting that the same scatter is shown by both $\epsilon_{\rm ff}$ and $\epsilon$ that, in the same
set of observations, agree within a factor of $\sim 2$ (\citealt{wu10,eva14,hey16,lee16}). 
These measures are performed with different methodologies, that involve tracers of both the stellar
\citep{eva14,hey16,vut16} and molecular gas components \citep{lad10, wu10,gol17}. 
The reasons for the wide scatter shown by these measures are highly debated and present a serious
challenge to modern SF theories (e. g., \citealt{hop14,gri19,seg24}).
These aim to explain the SFE of molecular clouds based on their turbulent properties, Mach number or 
virial parameter \citep{kru05,hen11,pad11,fed12} or balance between gravity and massive stellar feedback (e. g., \citealt{ras16,gru18}), 
or to the stochasticity of SF itself, that can account for this huge scatter only up to a limited extent \citep{gru19}.
It is also likely that such scatter is intrinsic to the molecular clouds and reflects 
a strong variation with time of the SFR and gas mass, 
occurring before the dispersal of the gas \citep{mur11,fel11}. 

At variance with local spirals, observations of star-forming regions in the local, compact starburst galaxy IRAS0
show a two orders-of-magnitude variation in $\epsilon_{\rm ff}$, reaching extreme values as high as 100 \% \citep{fis22} in the central region. 
This galaxy is located above the local SFR-M$_{\rm *}$ Main sequence where most galaxies lie, therefore, 
for its intense  star formation activity, IRAS0 is regarded as similar to the 
turbulent, compact starburst galaxies present at high redshift. For some of these systems, separate studies confirm 
$\epsilon_{\rm ff}$ values higher than local ones (e. g., \citealt{des23}). 

One major, currently unanswered question concerns the physical conditions leading to the formation of bound
star clusters, particularly in relation to the SFE.

From the theoretical point of view, a long-standing idea concerns a 'canonical' value
of $\epsilon = 50\%$ for the SFE, based on the virial theorem to explain the survival
of a bound cluster after mass loss on timescales less than one crossing time \citep{mat83}. 
However, in the case of slower gas loss, it is possible for a cluster to survive with lower SFE values  
(e. g., \citealt{boi02,far18}.)
Other studies have investigated the relation between $\epsilon$ and the bound fraction with
N-body simulations (e.g., \citealt{bau07,shu17}), 
showing that $\epsilon>30~\%$ is required to form gravitationally bound star clusters.

More works based on radiation hydrodynamics simulations showed that in high-density clouds
with surface density $\Sigma_{\rm gas}>10^2$M$_{\odot}$ pc$^{-2}$, 
the SFE and the bound fractions can 
incease significantly due to inefficient photoionisation feedback 
in a deep gravitational potential \citep{fuk21}.

From the observational point of view, the measure of $\epsilon$ is more problematic
in local young star clusters, in most cases because of the poor available constraints on
the amount of cold gas in the systems \citep{par09}. Exceptions are the bright centres
of a few nearby starbursts, that represent the most favourable systems where such measures can 
be performed. 
In a few cases, such sites are the hosts of super star clusters, where high SFE values are observed. 
Through the detection of the J= 3-2 rotational CO transition of CO in the local dwarf galaxy NGC 5253,
\cite{tur15} detected a young, massive (with M$_{\rm *}\sim 10^6$ M$_\odot$) star cluster with $\epsilon > 50$ \% efficiency.
Similar results are found for the SSCs of other systems such as NGC 253, NGC 4945 and Mrk 71A, 
which present SFE between 50 and 80 $\%$ \citep{ric20,emi20,oey17}. 
 
Other empirical, sometimes indirect arguments suggest that bound star clusters are likely characterised by high SFE.
In particular, arguments related to the dynamics of the gas in clusters, 
and the response of the surrounding envirnment to the energetic 
output from massive stars allow one to derive constraints on the SFE \citep{hil80}. 

In the Milky Way, one of the best studied YMCs is Westerlund I.  
At its young age (4.5-5 Myr, \citealt{cro06}) this system is expected to be out of virial equilibrium
because of a recent expulsion of residual gas not converted into stars \citep{cot12}.
One likely explanation for Westerlund I 
to survive the gas expulsion is a high star-formation efficiency, which would cause the cluster to remain close to virial equilibrium
\citep{men09,cot12}.

\cite{bas06} considered the luminosity profiles of a set of young clusters, with ages $<100$ Myr and, from the study
of their dynamical mass to observed light ratios, found that that several of them were out of virial equilibrium.
By means of N-body simulations of clusters including the effects of rapid gas loss,
they quantified the effect of rapid gas removal on the cluster disruption,
finding that models characterised by SFE between 40\% and 50\% best reproduced the observed dataset. 
\cite{hen12} studied the dynamical state of the Large Magellanic Cloud young massive cluster R136 through
a multi-epoch spectroscopic data analysis of a set of individual stars.
From the computed velocity dispersion, they concluded that 
R136 is in virial equilibrium and, comparing the low velocity 
dispersion with the values found in a few other young massive clusters, their result suggests that gas expulsion had 
a negligible effect on its dynamics. Among a few possibilities, this could be explained also with a 
high star-formation efficiency \citep{goo06}. \\

In cosmological simulations, an extensive test of the role of the SFE per freefall time
on galaxy and cluster properties was performed
by \cite{li18}. 
In this framework, star clusters are not resolved, but introduced in a sub-grid fashion, where 
star particles within molecular clouds are allowed to accrete gas from their surroundings.
Feedback from new stars can stop this process, and the final particle mass represents the one of a star cluster.  
While the global galaxy properties were weakly affected by the chosen value of
$\epsilon_{\rm ff}$ across a wide range, the cluster properties were rather sensitive to this parameter, in particular the
fraction of clustered star formation, showing that to reproduce the increasing observed trend of the cluster
formation efficiency with SFR, large values ($50 \% - 100 \%$) are required.
\cite{bro22} improved the study of \cite{li18}, with more comprehensive initial conditions describing more MW-mass
progenitors and run down to z=0. Among the explored set of values for $\epsilon_{\rm ff}$, only $\epsilon_{\rm ff}=1$ allows them to reproduce
the MW GCs mass function, whereas a too low efficiency of $\sim 1$ percent produces too few massive clusters and urealistic age spreads.

\cite{pol24} use high-resolution simulations of turbulent, isolated clouds of various masses  where feedback from individual stars is resolved. 
Their results support that the star formation of YMCs is rapid, with a large fraction (up to 85 \%)
converted into stars within the first freefall time of
the collapsing clouds. They stress that an inadequate treatment of feedback from individual stars might lead to overestimating the SFE. 

These results highlight the need for additional investigations, possibly within a cosmological framework, to address the interplay of the
SFE and stellar feedback, including more physical ingredients such as ionising radiation. 
In a forthcoming paper, we will study the impact of ionising feedback from individual stars on the self-regulation
of SF, in particular testing various values for the SFE. 

\section{Conclusions}
Star clusters represent the building blocks of galaxies and have recently
gained interest in galaxy formation.
Resolving the formation of star clusters 
stands out as one of the major ambitions in cosmological simulations, where, due
to significant technical challenges and computational limitations, they are often modelled as single particles, whose
properties are studied by means of sub-grid recipes. 
The study of their intenal properties in realistic galactic environments requires high-resolution simulations, 
most preferably at least at the sub-pc level.  
Clearly, even such high resolution is not enough to model the formation of single stars, but it is sufficient 
to quantify how stellar cluster substructures are affected by fast, small-scale
gravitational processes such as tidal shocks in the rapidly changing, primordial
density field, and to model properly the multi-scale, turbulent nature of star formation.
Building upon a previous project aimed at SImulating the Environment where Globular clusters Emerged (SIEGE, \citealt{cal22,rpas23}),
by means of cosmological simulations we studied the roles and the interplay of three fundamental factors, namely (i) stellar feedback,
(ii) star formation efficiency and (iii) stellar initial mass function, in shaping the intrinsic properties
of star clusters within a cosmological framework.
We compared the results obtained in the previous work of FC22, where the pre-SN feedback was modelled through the
instantaneous release of a significant quantity of mass and energy by massive stars at their birth, with another model in which
stellar winds are implemented, and where massive stars release mass and energy continuously at a low and constant rate. 
We considered two different models including this feedback scheme, but with two different values for the SFE,
expressed as a fraction of the free fall time, i.e. $\epsilon_{\rm ff}=0.1$ and $\epsilon_{\rm ff}=1$.
In one additional model, we matched the stellar winds prescriptions with a top-heavy stellar IMF, characterised
by the same mass interval as adopted in the standard case but significantly flatter slope, and therefore
more massive stars per stellar mass formed than with the fiducial initial mass function. 
All the simulations have been run down to $z=10.5$, suitable to appreciate the differences between the models and
for long enough to characterise the properties of the stellar systems.
In the present work, we have focused on the global properties of the simulations, of the structure at the centre of the simulated
box and on some star clusters features.  
In a forthcoming paper we will study in more detail some fundamental scaling relations of the stellar clusters and
their evolution on a longer timescale (Pascale et al., in prep). 

Our conclusions can be summarised as follows.

\begin{itemize}
\item While the large-scale properties of the gas are similar in all simulations, 
  the sub-structural properties of the star-forming clouds are sensitive to the different ingredients considered in our study.
  At the analysed redshifts, the gas clumps of the FC22 model show a smooth density field. 
  With its gentler release of energy and mass, the adoption of stellar winds is the ingredient making the biggest difference with
  respect to the old model, as it generates a more perturbed distribution, with the degree of
  complexity and filamentary structures increasing with decreasing redshift. 
\item This is also supported by the evolution of the gas density PDF. The PDF is narrow in the case of the FC22
  model, while it spans a larger range of gas densities  in the 'Winds' models.
  The three models with stellar wind prescriptions but different SFE and IMF do not show remarkable differences between each other
  in either the    density maps or the PDFs.
  \item In virialized systems like our clumps, the gravitational pressure $P_{\rm grav}$ and the
  1D density-weighted velocity dispersion.
  are quantities suited to measure the degree of
  turbulence. All our models show comparable values of $P_{\rm grav}$, although with different morphological distributions. 
  Similar to the gas density,   
  all the 'Winds' model show filamentary patterns, in particular at late times, regardless of the adopted SFE or IMF.  
 At $z=10.5$, all the 'Winds' models show very similar values for the 1D density-weighted velocity dispersion,
  a factor $\sim 2$ lower than the value computed for the FC22 model, again to be ascribed to a stronger pre-SN feedback of the latter. 
\item While massive stars start shaping the properties of the ISM immediately after their birth,
SNe may find different conditions in the gas in which they explode.   
In the FC22 model, many SNe explode in gas that had the time to recollapse after the prompt release of
energy from newly born massive stars. The 'Winds' models show generally a clear anticorrelation between T and $\rho$,
more similar to typical ISM  phase diagrams driven by stellar feedback.
The 'Winds' model with SFE=0.1 and 1.0 show significant clustering of the feedback sources,
that reflects the more compact stellar aggregates. 
\item
In comparison with the results of FC22 and adopting the same SFE of 10\%,    
the use of a more gentle pre-SN stellar feedback scheme, characterised by ejection of mass and energy via continuous stellar winds, 
produces a nearly 2 orders of magnitude increase in the maximum stellar density in the central region of the simulation at $z=10.5$. 
On the other hand, the 'Winds, SFE=1.0' model shows remarkably high stellar density already in the earliest phases, with values up to  
$10^4 M_{\odot}$ pc$^{-2}$ soon after the beginning of star formation. 
From the analysis of the PDF of the central stellar density across different models at the final redshift of our simulations,
the 'Winds, SFE=1.0' is the one that shows the distribution skewed towards the highest density values.
On the other hand, due to excessive stellar feedback related to the higher number of massive stars, the 'Winds, THIMF'
is the model characterised by the lowest density in stellar clumps.

\item We analysed the relation between stellar density ($\Sigma_{\rm *}$) and size for the clusters identified
in our models and compared it to observational datasets, incuding both  local star clusters and high-redshift clumps.
While the observed relation of local star clusters is nicely reproduced by the 'Winds, SFE=1.0', all the other models
fail in producing dense enough clusters. 
This confirms that the adoption of a high SF efficiency is the key ingredient that allows us to 
achieve realistic star clusters, characterised by properties comparable to the ones of local YSCs.

In the future, we will test new ingredients in our models. In particular, 
we will study the role of ionising radiation on the formation of star clusters and the implications regarding the other quantities
analysed here, such as the stellar densities in the most massive star clusters, that sometimes reach values in excess
of $10^5$ M$_{\odot}$ pc$^{-2}$. 
A follow-up study of early cluster formation in a more massive, Milky-Way mass halo will also be necessary.  
   
\end{itemize}

\begin{acknowledgements}
The research activities described in this paper have been co-funded by the European Union – NextGeneration EU within
PRIN 2022
project n.20229YBSAN - Globular clusters in cosmological simulations and in lensed fields: from their birth to the present epoch.
We acknowledge support from the 2023 INAF Mini-Grant ‘Clumps at cosmological distance: revealing their
formation, nature, and evolution' (1.05.23.04.01). 
This paper is supported by the
Fondazione ICSC, Spoke 3 Astrophysics and Cosmos Observations. National
Recovery and Resilience Plan (Piano Nazionale di Ripresa e Resilienza, PNRR)
Project ID CN\_00000013 “Italian Research Center for High-Performance Computing, Big Data and Quantum Computing”
funded by MUR Missione 4 Componente 2 Investimento 1.4: Potenziamento strutture di ricerca e creazione di
“campioni nazionali di R\&S (M4C2-19)” – Next Generation EU (NGEU). 
EL acknowledges financial support from the European Research Council for the ERC Consolidator grant DEMOBLACK, under contract no. 770017.
Enrico Vesperini acknowledges support from NSF grant AST-2009193 and from the John and A-Lan Reynolds Faculty Research Fund. 
We acknowledge EuroHPC JU for awarding the project IDs EHPC-REG-2021R0052 and EHPC-REG-2024R01-042 access to DISCOVERER at the Sofia Tech Park, Bulgaria.
This research was supported in part by Lilly Endowment, Inc., through its support for the Indiana 
University Pervasive Technology Institute. 

\end{acknowledgements}

\bibliographystyle{aa}

\begin{appendix}

\section{Star formation history of the central clump}
\label{sec_sfh}
\begin{figure}
  \includegraphics[width=9.2cm,height=6.92cm]{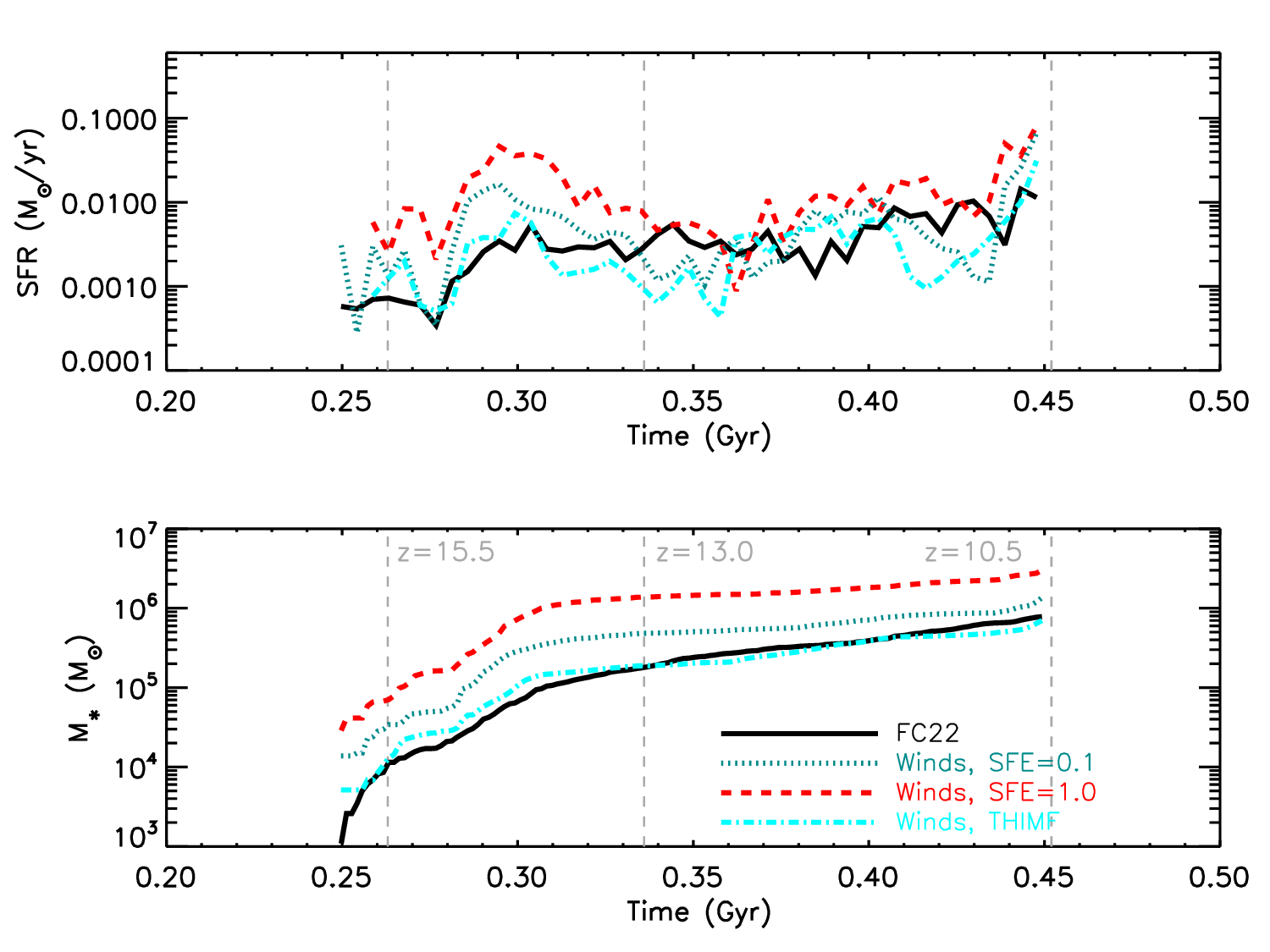}
\caption{Star formation history of the central regions of the box (see Fig.s \ref{fig_map_gas} and \ref{fig_map_stars}) 
  in the models considered in the present work. In the upper and lower panel we show the evolution of
  the SFR and the cumulative stellar mass, respectively, obtained in the 'FC22' (black solid line),
  'Winds, SFE=0.1' (dark-cyan solid line),  'Winds, SFE=1.0' (red solid line) and 'Winds, THIMF' (solid cyan line).   
  The three vertical dashed lines represent the cosmic times of the snapshots considered in this works, for
  which we report the corresponding redshifts. }
\label{fig_sfh}
\end{figure}

To better understand the differences in the features of our models, in this Section 
we analyse the SFHs calculated in the central regions of the box and shown
in Figs.  \ref{fig_map_gas}, \ref{fig_press_gas}, \ref{fig_map_sfr}, \ref{fig_map_stars}. 
In the upper and lower panels of Fig. ~\ref{fig_sfh} we show 
the evolution of the star formation rate and the cumulative stellar mass, respectively,
of the central regions of the box for the 4 models considered in this work. 
The model with the most massive central clump at $z=10.5$ is the 'Winds, SFE=1.0',
characterised by M$_* = 3 \times 10^6$ M$_{\odot}$.
By comparing the results of the 'Winds, SFE=0.1' and  'Winds, SFE=1.0' models, 
it is interesting to note that a factor 10 increase in the SFH results in a factor $\sim 2$ increase
in the final stellar mass at $z=10.5$.
Despite significant differences in both the feedback implementation  and adopted IMF,
another important aspect is the remarkable similarity between the cumulative stellar mass shown by the 
'FC22' and 'Winds, THIMF'. 
The final cumulative stellar mass of the 'Winds, SFE=1.0' model
is a factor $\sim 2$, 3.1 and 4.3 higher than the one of the 
'Winds, SFE=0.1', 'FC22' and 'Winds, THIMF', respectively.

\section{A few dense star clusters and clumps}
\label{sec_more_clumps}
\begin{figure*}
\center
\includegraphics[width=13.3cm,height=10cm]{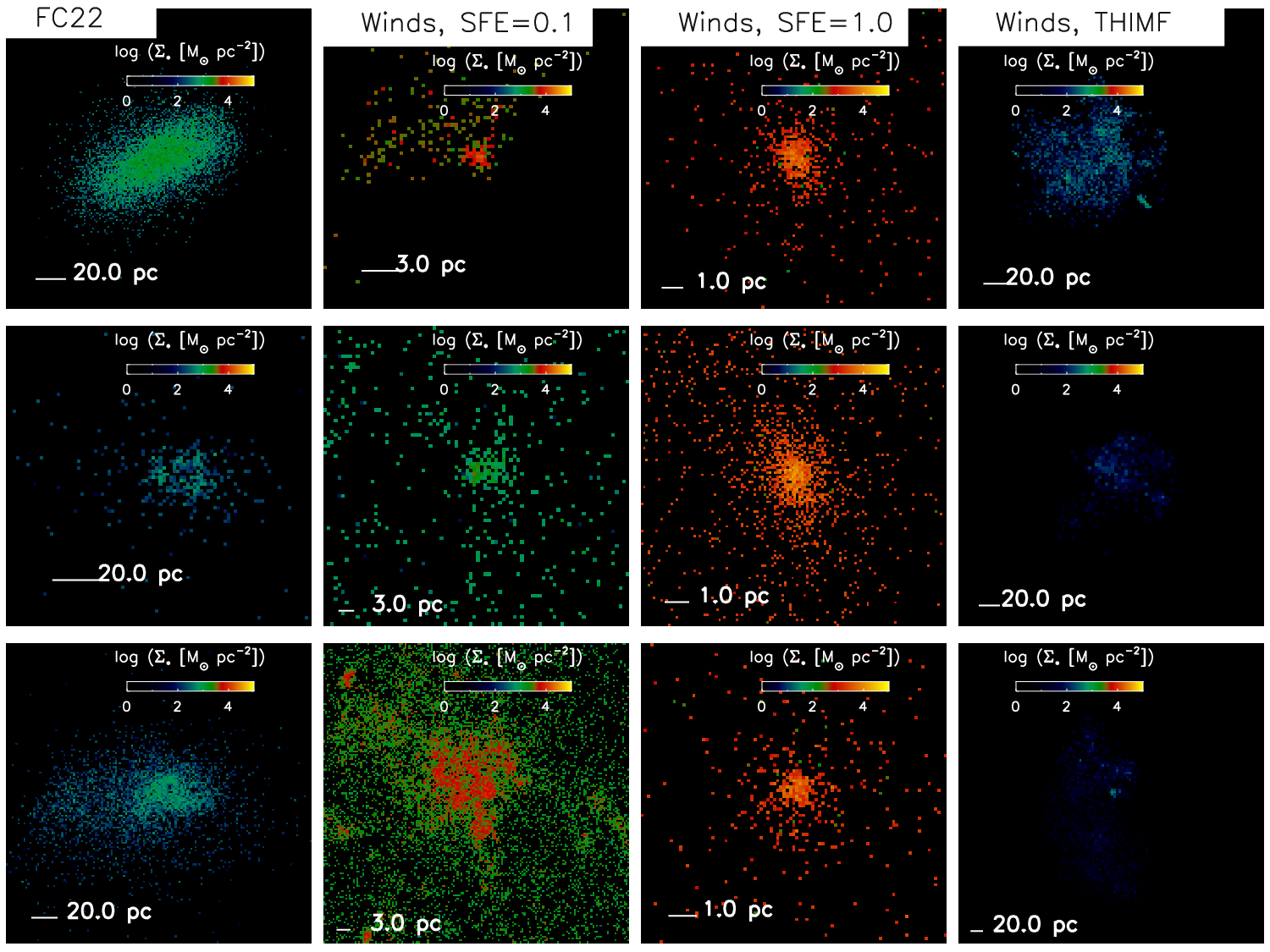}
\caption{Projected stellar density maps in the x-y plane for a few representative stellar clumps
  and star clusters in our models at $z=10.5$.
  Starting from the left,  in the first, second, third and fourth column we show clumps or clusters 
  in the 'FC22', ' Winds, SFE=0.1', 'Winds, SFE=1.0' and 'Winds, THIMF' model, respectively. 
  The horizontal white solid lines shown in each panel indicates the physical scale.} 
\label{fig_map_clumps}
\end{figure*}

In Fig. ~\ref{fig_map_clumps} we show the stellar density maps for some representative clumps and clusters of each model. 
This figure highlights further the different features of the stellar aggregates created in our
simulations. 
Very diffuse, extended clumps are formed within the 'FC22' model (first colums from left of Fig. ~\ref{fig_map_clumps}). 
The change of prescriptions from the 'FC22' model, characterised by instantaneous and intense release of energy
and mass in the pre-SN phase, 
to the low-intensity stellar winds of the 'Winds, SFE=0.1' model
leads to a major transition from extended clumps 
to very compact star clusters, with maximum densities larger by one order of magnitude or more (second column). 
By adopting a factor 10 higher SFE efficiency, in the 'Winds, SFE=1.0' model the cluster densities
may reach values up to $10^4~M_{\odot}$ pc$^{-2}$ and sizes of the order of $\sim 1$ pc (third column).
Finally, again due to the stronger feedback, the clumps of the 'Winds, THIMF' model  (fourth column) 
are characterised by extended, scattered stellar distributions with compact 
knots at their centre, but presenting significantly lower maximum densities, of $\Sigma_{\rm *}\sim 10^2~M_{\odot}$ pc$^{-2}$.

\end{appendix}






%




%



\end{document}